%% file: main.tex
\pdfoutput=1
\newif\ifarxiv 
\newif\ifcamera
\arxivtrue
\camerafalse
\newif\ifdeanon
\ifarxiv\deanontrue\fi
\ifcamera\deanontrue\fi

\documentclass{article}

\ifarxiv
  \usepackage[preprint]{neurips_2026}
\else\ifcamera
  \usepackage[eandd,final]{neurips_2026}
\else
  \usepackage[eandd]{neurips_2026}
\fi\fi

\usepackage[utf8]{inputenc}
\usepackage[T1]{fontenc}
\usepackage[dvipsnames]{xcolor}
\ifarxiv
  \usepackage[unicode,bookmarks=false,breaklinks=true]{hyperref}
  \hypersetup{colorlinks=true,citecolor=RoyalBlue,linkcolor=RoyalBlue,urlcolor=RoyalBlue}
\else
  \usepackage[unicode,bookmarks=false,breaklinks=true,hidelinks]{hyperref}
  \hypersetup{citecolor=black,linkcolor=black,urlcolor=black}
\fi
\usepackage{url}
\usepackage{booktabs}
\usepackage{amsfonts}
\usepackage{amsmath}
\usepackage{amssymb}
\usepackage{nicefrac}
\usepackage{microtype}
\usepackage{graphicx}
\usepackage{subcaption}
\usepackage{multirow}
\usepackage{enumitem}
\usepackage{algorithm}
\usepackage{algorithmic}
\usepackage{listings}
\usepackage{xspace}
\usepackage{wrapfig}
\ifarxiv
  \newcommand{\pbtmintedsetup}[1]{%
    \def\pbtmintedlang{#1}\def\pbtmintedtext{text}%
    \ifx\pbtmintedlang\pbtmintedtext
      \lstset{language={},basicstyle=\ttfamily\scriptsize,numbers=none,frame=none,
              xleftmargin=0pt,breaklines=true,columns=fullflexible,keepspaces=true}%
    \else
      \lstset{language=Python,basicstyle=\ttfamily\small,numbers=left,columns=fullflexible,keepspaces=true}%
    \fi}
  \lstnewenvironment{minted}[2][]{\pbtmintedsetup{#2}}{}
\else
  \usepackage{minted}
\fi
\usepackage[normalem]{ulem}

\newcommand{\benchname}{PBT-Bench\xspace}
\newcommand{\numproblems}{100\xspace}
\newcommand{\numlibraries}{40\xspace}

\definecolor{darkgreen}{RGB}{0,120,0}

\title{\benchname: Benchmarking AI Agents on Property-Based Testing}

\author{%
\ifdeanon
  Lucas Jing\thanks{Equal contribution.}%
  \textsuperscript{,}%
  \thanks{Work done during an internship at the University of Washington.} \\
  Tsinghua University \\
  jgh23@mails.tsinghua.edu.cn
  \And
  Xinqi Wang\footnotemark[1]  \\
  University of Washington \\
  wxqkaxdd@cs.washington.edu \And
  Liao Zhang \\
  Liaoyuan Technology \\
  zhangliao714@gmail.com
  \And 
  Simon S. Du \\
  University of Washington \\ 
  ssdu@cs.washington.edu
\else
  Anonymous Author(s)
\fi
}

\begin{document}

\maketitle

\begin{abstract}
\input{sections/abstract}
\end{abstract}

\input{sections/introduction}
\input{sections/related_work}
\input{sections/benchmark}
\input{sections/experimental_setup}
\input{sections/results}
\input{sections/analysis}
\input{sections/conclusion}

\ifdeanon
  \begin{ack}
  \input{sections/acknowledgments}
  \end{ack}
\fi

\bibliographystyle{plainnat}
\bibliography{references}

\appendix
\input{sections/appendix}

\ifarxiv\else
  \newpage
  \input{checklist}
\fi

\end{document}

%% file: sections/abstract.tex
Existing code benchmarks measure whether an agent can produce \emph{any} test that reproduces a known bug, or whether it can produce a patch that fixes a
described issue.
Neither isolates the distinct skill of \emph{property-based testing}:
reading documentation to identify a rule the code must always obey, and then
writing a random-input generator targeted enough to find an input that breaks it.
We introduce \emph{\benchname}, a benchmark of \numproblems\
curated property-based testing problems across \numlibraries\ real Python
libraries.
Each problem injects one or more semantic bugs (365 in total, mean 3.65 per problem) designed so that
generic random inputs almost never trigger them; the agent must read
the library's documentation, identify the relevant invariant, and write a
Hypothesis property test whose random inputs include the rare cases that trigger the bug.
Bugs are stratified across three difficulty levels (L1--L3), from
single-constraint boundary bugs to stateful, cross-function protocol
violations.
We evaluate eight contemporary LLMs under two prompting regimes
(open-ended baseline vs.\ explicit Hypothesis scaffolding) for three
independent runs per configuration, and additionally evaluate the frontier
model Claude Opus~4.8 on the same scaffold and inside Claude Code.
Bug recall under the PBT-guided prompt ranges from 42.1\% to 83.4\% across
models; Opus~4.8 reaches 81.2\% and does not exceed the best original model, and two
bugs are never found in two of three runs by any model.
The gain from the PBT prompt varies from $-8$ to $+42$ percentage points and is not
monotone in model strength: Opus~4.8 writes strong property tests when asked, but
under the open-ended prompt it stops after exposing a single bug.
A controlled ablation shows that for Opus~4.8 the framework instructions (use
Hypothesis and test the whole documented specification) account for essentially
all of the gain, whereas the property taxonomy we provide adds recall for
mid-capability models.
The hardest bugs are model-specific, leaving gaps that no single model closes.
We release the benchmark, harness, and full evaluation corpus
to support downstream work on documentation-grounded semantic
reasoning.

%% file: sections/introduction.tex
\section{Introduction}
\label{sec:introduction}

Modern language models pass most code generation benchmarks when the expected
behavior is stated concretely through input--output pairs. They do less well
when the specification is \emph{abstract}: a documented invariant that must
hold across an entire input distribution rather than on any particular
example.
Property-based testing (PBT), introduced by QuickCheck~\citep{claessen2000quickcheck}
and popularized in Python by Hypothesis~\citep{maciver2019hypothesis}, is a
testing methodology in which a programmer specifies (i)~a \emph{property} --- a
predicate that should hold universally --- and (ii)~a \emph{strategy} --- a
typed, composable random-input generator --- for sampling from the intended
input distribution.
The framework then searches for inputs that violate the property and shrinks
any counterexample to a minimal failing form.
A primer on Hypothesis strategies is in Appendix~\ref{app:hypothesis-primer}.
Figure~\ref{fig:pbt-qsort-bug} shows a minimal example: a quicksort that drops elements equal to the pivot violates the properties ``output is sorted and a permutation of the input'' on random integer lists that contain duplicates.

PBT provides greater coverage through random input exploration and expresses specifications as machine-checked, living documentation~\citep{ravi2025empirical}.
However, writing a good property test is harder than a unit test~\citep{ravi2025empirical, goldstein2024property}: the author must abstract from concrete outputs to invariants and design an input distribution that targets likely violations---a barrier that LLMs are naturally positioned to lower.

Recent benchmarks have begun to probe LLMs' testing abilities, but none
isolates the PBT skill. \textsc{SWE-bench}~\citep{jimenez2024swebench} rewards
\emph{patch synthesis} given a bug report; \textsc{SWT-Bench}~\citep{muendler2024swtbench}
rewards \emph{test synthesis} given a bug report; \textsc{TestExplora}~\citep{testexplora2026}
rewards test synthesis from documentation, but over a mined distribution of
real-world pull requests that is dominated by bugs whose fixes include
dedicated unit tests, I/O coupling, or integration-level failures; and
mutation-based evaluations score LLM-written tests against automatically
generated mutants~\citep{vikram2023pbtllm,sun2026swemutation}.
These settings either accept finite, concrete test cases or rely on mechanically
generated faults. None evaluates agents on the step that PBT requires, turning
an abstract specification into universal properties and programmatic input
distributions, against faults designed so that only such properties expose them.

\begin{figure}[t]
\centering
\begin{minted}[fontsize=\small, linenos]{python}
from hypothesis import given, strategies as st

def qsort(xs):
    if not xs: return []
    p = xs[0]  # pivot element at each recursion step
    return qsort([x for x in xs[1:] if x < p]) + [p] + \
           qsort([x for x in xs[1:] if x > p])  # BUG: ignores elements equal to pivot

@given(st.lists(st.integers()))
# Built-in strategy generating arbitrary integer lists
def test(xs):
    ys = qsort(xs)
    # properties: sorted output + multiset equivalence with input
    assert is_sorted(ys) and same_multiset(xs, ys)
\end{minted}

\caption{A Hypothesis property test exposing a quicksort bug. The \texttt{@given} decorator binds a strategy (here random integer lists) to the test's argument; lists with duplicates trigger the missing-pivot defect, caught by the sortedness and permutation properties, and Hypothesis shrinks the failing input to a minimal one.}
\label{fig:pbt-qsort-bug}
\end{figure}

We argue that evaluating PBT as a distinct skill requires
curated --- not mined --- bugs, because not every bug is PBT-testable. A bug is
PBT-testable only if (a)~a semantic invariant of the API is violated,
(b)~the violation manifests under some expressible Hypothesis strategy without
environmental mocks, and (c)~the violation is deterministic under random input
sampling. In practice, many real-world pull request bugs do not satisfy at least one
of these conditions --- common examples include UI rendering failures, database
state dependencies, network ordering effects, and type errors that lack a
corresponding documented invariant.
A benchmark mined from PR history therefore conflates PBT skill
with the agent's ability to navigate around non-PBT-testable cases.

\paragraph{Contributions.}
\begin{enumerate}[leftmargin=*, topsep=2pt, itemsep=2pt]
  \item \benchname, a curated, documentation-grounded benchmark of \numproblems\
  property-based testing problems across \numlibraries\ real Python libraries
  (serialization, data structures, date--time, numerics, type systems, state
  machines, parsing), each designed and verified to satisfy the PBT-testability
  conditions above. Every bug comes with a ground-truth Hypothesis strategy, a
  precise trigger condition, and a difficulty label (L1--L3). Unlike mutant-based
  evaluations of LLM-written tests (Section~\ref{sec:related_work}), the faults are
  LLM-proposed and author-reviewed, selected to escape the library's own tests,
  and must be exposed by documentation-derived properties.
  \item An automated, containerized F$\to$P (fail on buggy, pass on fixed)
  harness that scores each test function separately for each injected bug, so
  that one test file can be credited for several bugs.
  \item A large-scale evaluation: eight contemporary LLMs (Claude Sonnet 4.6,
  DeepSeek V3.2, Gemini 3 Flash, GLM 5.1, Grok 4.1 Fast,
  Qwen 3.6 Plus, Qwen3.5-35B-A3B, Step 3.5 Flash) in two prompting regimes
  (open-ended baseline and explicit Hypothesis scaffolding), three runs each
  ($4{,}800$ trajectories), plus a frontier model (Claude Opus~4.8) on the same
  scaffold and in Claude Code, and a controlled prompt ablation.
  \item Findings: (i)~the benchmark is not saturated: the best agent reaches
  83\% bug recall, Opus~4.8 does not exceed it, and two bugs are never found in
  two of three runs by any model; (ii)~the gain from the PBT prompt ranges from
  $-8$ to $+42$ percentage points and is not monotone in model strength; for
  Opus~4.8 it comes from the framework instructions (use Hypothesis, cover the
  specification), while the property taxonomy adds recall for mid-capability
  models; (iii)~no single model dominates the hardest bugs, and the union of all
  sixteen original model--prompt cells reliably finds 99.5\% of bugs, 12.7 points
  above the best single cell.
\end{enumerate}

We release the benchmark, harness, problem templates, and full evaluation
corpus (code under MIT, data under CC~BY~4.0) at
\url{https://github.com/ElliotXinqiWang/PBTbench} and
\url{https://huggingface.co/datasets/pbtbench-team/pbt-bench}.
All problem files contain the \textsc{BIG-bench} canary string to aid
decontamination of future training corpora~\citep{srivastava2023bigbench}.

%% file: sections/related_work.tex
\section{Related work}
\label{sec:related_work}

\paragraph{Benchmarks for patch and test synthesis.}
\textsc{SWE-bench}~\citep{jimenez2024swebench} established the template for
coding-agent evaluation: given an issue and a repository snapshot, produce a
patch that passes a held-out test. \textsc{SWE-bench Verified}~\citep{swebenchverified2024}
curates a higher-signal subset, and later variants
(\textsc{SWE-bench Multimodal}~\citep{yang2025swebenchmm},
\textsc{SWE-Lancer}~\citep{miserendino2025swelancer},
\textsc{SWE-EVO}~\citep{swebenchevo2025}) extend scope and difficulty.
\benchname\ does \emph{not} ask the agent to fix the bug, only to \emph{expose} it;
repair and exposure are known-distinct skills~\citep{muendler2024swtbench}.
\textsc{SWT-Bench}~\citep{muendler2024swtbench} re-frames SWE-bench for test
generation with the fail-to-pass (F$\to$P) criterion that underlies our
per-function scoring; it builds on mined bug corpora such as
\textsc{Defects4J}~\citep{just2014defects4j} and \textsc{BugsInPy}~\citep{widyasari2020bugsinpy}.
\benchname\ differs in two ways: the agent never sees a bug description (the
library's documentation is the only oracle), and the test must be a Hypothesis
\texttt{@given} property rather than a concrete assertion that could encode the
fixed behaviour's output as a literal.
\textsc{TestExplora}~\citep{testexplora2026}, closest in framing, mines 2,389
documentation-as-oracle tasks from pull requests; its natural bug distribution
includes many concrete-assertion-testable and I/O-coupled tasks, whereas
\benchname\ isolates the PBT skill on a controlled distribution.
Code-synthesis benchmarks (\textsc{HumanEval}~\citep{chen2021humaneval},
\textsc{LiveCodeBench}~\citep{jain2025livecodebench}, \textsc{ClassEval}~\citep{du2024classeval},
\textsc{AutoCodeBench}~\citep{chou2025autocodebench}) do not target testing; the
construction protocol of \textsc{Terminal-Bench}~\citep{merrill2026terminalbench}
informs our QA pipeline (Section~\ref{sec:benchmark}).

\paragraph{Mutation testing and mutant-based evaluation of LLM tests.}
\benchname\ is structurally a mutation-style evaluation: seeding faults and
counting how many a test suite exposes is the core of mutation
testing~\citep{jia2011mutation,papadakis2019mutation}, and mutant detection is
correlated with real-fault detection~\citep{just2014mutants}. Mutants are
already used to evaluate or guide LLM test generation:
\citet{vikram2023pbtllm} score LLM-written property tests against mutants,
\citet{dakhel2024mutation} and \citet{wang2026mutation} feed surviving mutants
back to the model, \citet{harman2025mutation} deploy mutation-guided test
generation at Meta, and \textsc{SWE-Mutation}~\citep{sun2026swemutation} measures
the mutation score of LLM-generated test suites for real repositories. We do not
claim a new evaluation paradigm. \benchname's contribution is narrower: a curated,
documentation-grounded, Hypothesis-constrained benchmark for PBT by coding agents.
Many of its faults resemble mutation operators in form; they differ in being
LLM-proposed and author-reviewed, selected (not generated) for PBT-testability,
survival of the library's own test suite, depth below the API, and difficulty,
with several per problem, each triggered by different inputs. The agent never sees
the faults, and the oracle is the documentation: we measure test generation from a
specification, not the adequacy of an existing suite.

\paragraph{Property-based testing and LLMs.}
PBT originated with QuickCheck~\citep{claessen2000quickcheck} and was ported to
Python as Hypothesis~\citep{maciver2019hypothesis}; it is used in industry to
find edge-case bugs~\citep{ravi2025empirical,goldstein2024property}.
LLM-based concrete test generators (\textsc{ChatUniTest}~\citep{chen2023chatunitest},
\textsc{CoverUp}~\citep{pizzorno2024coverup}, \textsc{CodaMosa}~\citep{lemieux2023codamosa},
\textsc{LIBRO}~\citep{kang2023libro}) produce input--output pairs, and LLM-driven
fuzzers (\textsc{TitanFuzz}~\citep{deng2023titanfuzz}, \textsc{FuzzGPT}~\citep{deng2024fuzzgpt})
generate programs without falsifiable predicates; PBT instead requires a universal
invariant paired with a typed input strategy.
\citet{properysolver2025} validate generated code against PBT invariants, and
\citet{xiong2026natural} derive executable properties for Android apps~\citep{xiong2024general}
with a staged LLM workflow.
Closest in spirit, \citet{agenticpbt2025} build an agent that infers properties
from code and documentation, writes Hypothesis tests, and triages its own
failures; across 100 popular Python packages, 56\% of its reports were valid bugs
on manual review, and three patches were merged upstream. That work shows that
agentic PBT finds real bugs in the wild; \benchname\ answers the complementary
measurement question of how well different agents do PBT relative to each other
and to ground truth. Because every bug has a known location, trigger region, and
reference strategy, misses are observable and recall is measurable, whereas in the
wild only precision can be assessed, by manual triage, and targets drift as bugs
are fixed; fixed problems, prompts, and scoring support repeated comparison. Conversely, a controlled
benchmark cannot capture real maintainer outcomes.

%% file: sections/benchmark.tex
\section{The \benchname\ Benchmark}
\label{sec:benchmark}

\benchname\ contains \numproblems\ curated property-based testing
problems across \numlibraries\ Python libraries. This section specifies the
problem format (\S\ref{subsec:format}), the three-level difficulty taxonomy
(\S\ref{subsec:difficulty}), the curation and quality-assurance protocol
(\S\ref{subsec:curation}), and the fidelity argument for injected bugs
(\S\ref{subsec:fidelity}).

\subsection{Problem format}
\label{subsec:format}



Each problem contains a buggy library, agent-accessible documentation (\texttt{docs/}; the sole oracle), a passing existing test suite, and one or more reversible bug patches (\texttt{bug\_N.patch}).
Reversibility is key: the default container state applies all patches; removing patch $N$ yields the fixed-bug-$N$ configuration, enabling independent per-bug F$\to$P scoring even when a problem contains multiple co-injected bugs.
Full directory layout and metadata schema are in Appendix~\ref{app:problem-format}.



\paragraph{PBT-testability criteria.}
Every injected bug must satisfy four criteria, which are enforced during curation and re-verified by \texttt{check\_infra.py}.
Specifically, we require \textit{semantic}, meaning the bug violates a documented invariant or contract rather than an implementation detail, including properties such as roundtrip identity, commutativity, monotonicity, bounded output ranges, and protocol preconditions; \textit{expressivity}, meaning the invariant can be encoded as a \texttt{@given} property using Hypothesis's built-in and \texttt{@composite} strategies without relying on external effects such as files, sockets, or clocks; \textit{stealth}, meaning the bug is not easily detectable by short manual inspection and typically resides in under-tested code paths or is masked by locally plausible computations; and \textit{deterministic trigger}, meaning there exists a well-defined input region $T$ on which the bug manifests with probability one (or a specific action sequence in the stateful case), with the ground-truth strategy concentrating probability mass on $T$.

\paragraph{Two-container evaluation harness.}
The benchmark is run in two independent containers per problem (Docker or
Apptainer, depending on the host environment). The
first (\emph{agent container}) contains the buggy library, documentation, and
existing test suite, and is handed to the agent under the OpenHands SDK~\citep{wang2025openhands} with
the prompt template of the relevant mode. The second (\emph{eval container})
is created after the agent writes \texttt{pbt\_test.py}: for each injected bug
$b$, the harness restores the buggy library, runs each test function, then
reverses only patch $b$ and re-runs. A bug is \emph{found} when at least one
test function passes the F$\to$P predicate --- fails on the buggy version and
passes on the version with $b$ reversed --- for that specific bug. This
per-function, per-bug scoring correctly credits agents who write one
test per bug in a single file.

\subsection{Difficulty taxonomy}
\label{subsec:difficulty}

Each injected bug is labelled with one of three difficulty levels reflecting
how sophisticated the input generator must be for random search to hit the
trigger region within 200 examples (the Hypothesis default; Appendix~\ref{app:hypothesis-primer} gives a primer).
\textbf{L1} corresponds to single-constraint bugs, where either Hypothesis's default strategies suffice (e.g., boundary exploration triggers the bug with high probability) or the agent needs to identify a single input constraint such as a specific value range, size class, or structural property; the trigger probability under default strategies ranges from high to below one percent.
\textbf{L2} denotes multi-constraint triggers that require simultaneously satisfying several documentation-derived conditions, such as value ranges, structural constraints, and operation ordering.
\textbf{L3} captures cross-function or cross-operation protocol violations, which require tests over sequences of API calls or metamorphic tests (i.e., checking invariants across multiple executions), where the invariant spans multiple API calls.


Across the \numproblems\ canonical evaluation problems, the 365 injected
bugs distribute as follows: L1 = 87 ($24\%$), L2 = 184 ($50\%$),
L3 = 94 ($26\%$).
The distribution is intentionally skewed toward L2--L3: early curation
rounds consistently found that trivial single-constraint bugs were solved
by both reference models in a single rollout, providing no discriminative signal;
most were retired or relabelled.

\subsection{Curation and quality assurance}
\label{subsec:curation}

\paragraph{Library selection.}
Libraries were selected for rich internal-state invariants not directly observable through the public API, a substantial gap between the public interface and its implementation (making code-reading localization hard), documentation quality sufficient for doc-first trigger inference, and pure Python implementation.
We excluded encode/decode core paths (roundtrip properties trivially cover their bugs) and libraries where patches are likely to conflict with internal self-use.

\paragraph{Bug design and retirement.}
Problems were developed through an iterative design-evaluate-retire process
in which an LLM-assisted design agent proposed candidate bugs and patches,
and human authors reviewed each candidate for correctness, stealth, and
PBT-testability before acceptance.
After each design round, every candidate problem was evaluated against two
\emph{reference models} (Claude Sonnet~4.6 and GLM~5.1, each running the
OpenHands baseline scaffold for a single rollout); problems on which both
reference models achieved full recall were retired or redesigned, as unanimous
success on a single rollout provides no discriminative signal.
The difficulty bar was raised progressively across rounds as earlier problems
proved easier than anticipated.
Retired problems are archived in the repository as documented negative
examples for future benchmark construction.

Each bug candidate was checked against a design checklist before acceptance:
injection at least two call-chain layers below the public API surface;
domain knowledge required to understand why the code is incorrect
(the bug must not be obviously wrong from a local reading);
the \emph{clean-pass criterion} (the injected bug must leave the library's existing unit test suite intact);
for problems containing multiple bugs, each bug must be independently triggerable
by a distinct input region that does not simultaneously trigger other bugs in
the same problem (verified via pairwise testing of ground-truth strategies);
and the existence of a ground-truth strategy derivable from documentation
alone that finds the bug within the default example budget.

\paragraph{Three-stage QA pipeline.}
Each accepted problem passed three verification stages.
\emph{Stage~1 (automated)} ran \texttt{check\_infra.py} to verify clean
patch application, upstream-test pass, trigger-path independence, and
ground-truth solution validity.
\emph{Stage~2 (reference-model gate)} evaluated each problem under the two
reference models defined above; problems where both achieved full recall
were retired.
\emph{Stage~3 (manual adversarial check)} confirmed that each accepted bug
satisfies the stealth criterion and that existing-test pass rates are unaffected.
The Stage~3 protocol is the same one applied in
Section~\ref{subsec:adversarial}.


\subsection{Fidelity and construct validity}
\label{subsec:fidelity}

Controlled bug injection raises a representativeness concern. Our position:
\emph{PBT-testability is a design constraint, not a distributional
one} --- a benchmark evaluating PBT must select for bugs PBT could in
principle detect. Within that constraint we preserve fidelity three
ways. First, every injected bug falls into one of eleven classes empirically observed in open-source Python bug reports (off-by-one/boundary, wrong reference/expression, condition/logic inversion, wrong operator, argument/field order swap, missing operation, traversal/ordering, sign/direction error, protocol/domain-specific, wrong constant/value, and state corruption/mutation); the full taxonomy with counts is in Appendix~\ref{app:bug-class-map}. Second, every bug
satisfies the \emph{clean-pass criterion} (defined in Section~\ref{subsec:curation}),
matching the empirical fact that released semantic bugs escaped their own tests.
Third, the \numlibraries\ libraries span seven domains (serialization, data
structures, date--time, type systems and schemas, numerics, state
machines, parsing/HTML); the full library list is in
Appendix~\ref{app:library-list}.

\paragraph{Construct validity.}
The targeted properties go beyond round trips, which account for only 18\% of bugs
by the \texttt{prop:*} tags in \texttt{problem.yaml}; the rest are structural
invariants, conformance to documented semantics, ordering, stateful, model-based,
and metamorphic properties. For four bugs in three problems we verified fixed
upstream counterparts, one of them found by a property-based test; one injected
bug modifies the very line that fixed a tomlkit issue and re-breaks that issue's
reproducer (Appendix~\ref{app:analogues}). These cases are illustrative, not a
systematic audit, and success on \benchname\ is not evidence of in-the-wild bug
finding, which \citet{agenticpbt2025} measure directly.

%% file: sections/experimental_setup.tex
\section{Experimental setup}
\label{sec:experimental_setup}

We evaluate eight contemporary LLMs under two prompting regimes, each repeated
three times per problem, over the \numproblems-problem evaluation set. All
configurations use the same OpenHands agent scaffold to isolate prompt and model effects from scaffold-specific tuning.
After review we added a frontier model, a commercial agent scaffold, and a
controlled prompt ablation (end of this section).

\paragraph{Models.}
We evaluate Claude Sonnet 4.6, DeepSeek V3.2, Gemini 3 Flash, GLM 5.1, Grok
4.1 Fast, Qwen 3.6 Plus, Qwen3.5-35B-A3B (Mixture-of-Experts, 3B active
parameters), and Step 3.5 Flash. All models are queried through OpenRouter
with the provider's default temperature. Model release dates span November 2025
through April 2026.

\paragraph{Prompt regimes.}
Each model is evaluated under two prompts:

\begin{itemize}[leftmargin=*, topsep=2pt, itemsep=2pt]
  \item \textbf{Baseline.} An open-ended four-phase instruction --- understand
  the library, find uncovered behavior, write a failing test, finalize --- that
  does not mention property-based testing, Hypothesis, or \texttt{@given}. The
  agent is told only that a test must fail on the current library and pass
  after the bug is fixed. This measures general testing ability.
  \item \textbf{PBT.} An explicitly structured four-step instruction that
  names Hypothesis as the testing framework, requires \texttt{@given}
  decorators, and teaches the Roundtrip/Model-oracle/Invariant property
  taxonomy. The agent is instructed to write at least five focused test
  functions, each covering one documented invariant, without first hunting
  for a specific bug. This measures PBT ability under scaffolding.
\end{itemize}

The prompt templates are reproduced in Appendix~\ref{app:prompts}.
Besides the prompt, the two regimes differ in that PBT containers preinstall
Hypothesis; the harness also sets a longer pytest timeout for PBT, but it never
binds (Section~\ref{subsec:ablation}). Section~\ref{subsec:ablation} separates the
framework instructions, which include Hypothesis availability, from the property
taxonomy and template.
No Baseline trajectory installed or imported Hypothesis, so Baseline tests are
concrete example-based tests.

\paragraph{Agent scaffold.}
All models run inside the OpenHands SDK v1.11.5
with the standard tool set (file editor, terminal, think) and a maximum of
200 tool-call iterations. Agent containers are built from
\texttt{python:3.12-slim} with the target library preinstalled via a cached
image layer per library. PBT-mode containers additionally preinstall
Hypothesis; Baseline containers do not, though the agent may install
it during its run. Each agent has 60~minutes of wall-clock time per problem.

\paragraph{Evaluation.}
After the agent writes \texttt{/workspace/pbt\_test.py}, the harness
collects the set of test functions, then for each injected bug
$b_i$: (i)~restores the buggy library from a pristine backup,
(ii)~runs each test function individually with a 30\,s limit (a test still running then counts as failing; Appendix~\ref{app:post-review}),
(iii)~reverses patch $b_i$, (iv)~re-runs each test function. A function
is \emph{useful for $b_i$} if it fails on the buggy version and passes
after $b_i$ is reversed. The bug is \emph{found} if at least one function
is useful for it. We report three metrics:
(1)~\textbf{bug recall}, $\mathrm{bugs\_found}/\mathrm{bugs\_total}$ averaged
over problems; (2)~\textbf{problem coverage} (F$\geq$1), the fraction of problems
with at least one bug found, measuring breadth; and (3)~\textbf{full recall}
(Perfect), the fraction of problems with all bugs found, measuring depth.
The released per-trajectory records additionally include test precision
(useful / total test functions).

\paragraph{How to use \benchname.}
The leading metric is \emph{PBT-mode bug recall} over three runs, with agents
compared by a paired bootstrap over problems (Appendix~\ref{app:stability}): it
measures the skill the benchmark isolates, turning documented invariants into
Hypothesis properties and strategies that expose semantic bugs. Full recall is
the most informative secondary metric, separating agents that find one bug per
problem from agents that test the specification exhaustively. The Baseline mode
and $\Delta=\text{PBT}-\text{Baseline}$ are diagnostics of how much an agent's
testing depends on being told to use PBT; they confound PBT skill with how
thoroughly the agent answers an open-ended request (Section~\ref{subsec:headline}),
so we recommend reporting them alongside, not ranking agents by them.

\paragraph{Post-review evaluations.}
\emph{Frontier model:} Claude Opus~4.8 on the same OpenHands scaffold with the
provider's default reasoning settings: four PBT, two Baseline, and three
framework-only runs (Appendix~\ref{app:post-review}).
\emph{Commercial scaffold:} Claude Code, the vendor's CLI agent, run headless with
Opus~4.8 and its default settings, the unchanged PBT prompt, network egress blocked except
for the model endpoint, a 15-minute limit (reached on 10 of 300 trajectories), and the
same F$\to$P scoring (three runs).
\emph{Controlled ablation:} a \emph{framework-only} prompt keeps the PBT prompt's
framework instructions (Hypothesis preinstalled, \texttt{@given} on every test,
\texttt{max\_examples=200}, tests derived from the documentation, ``cover the
specification rather than hunt for one bug'') but drops the property taxonomy, the
``5+ test functions'' target, and the template; it runs for Opus~4.8, Qwen~3.6 Plus,
Sonnet~4.6, and DeepSeek~V3.2, plus two paraphrases of it for Qwen~3.6 Plus.
A harness defect (since fixed) let one harness process stop another's agents; the
0--22\% of post-review trajectories per cell it ended are excluded, and all
post-review instances were re-scored (Appendix~\ref{app:post-review}). Our Codex CLI (GPT-5.5) integration
produced no test file in three pilot runs, so Codex is not reported.

\paragraph{Reproducibility and decontamination.}
All 4,800 trajectories, generated \texttt{pbt\_test.py} files, and container build scripts are released under a permissive licence, together with the post-review runs; one trajectory timed out and is recorded as not-found (Appendix~\ref{app:eval-timeout}).
Every problem file embeds the \textsc{BIG-bench} canary GUID~\citep{srivastava2023bigbench} for training-corpus decontamination.
Computational details and API costs are in Appendix~\ref{app:compute}.

%% file: sections/results.tex
\section{Results}
\label{sec:results}

We report results from 4,800 agent trajectories: eight models, two prompt
regimes, three independent runs per cell, \numproblems\ problems per run,
plus the post-review frontier evaluations of Section~\ref{sec:experimental_setup}.
Unless otherwise stated, numbers are averaged across runs.

\subsection{Headline results}
\label{subsec:headline}

Table~\ref{tab:headline} reports the primary metrics separately for the PBT
mode (panel a, the benchmark's leaderboard) and the Baseline mode (panel b, a
diagnostic). Three observations stand out.

\begin{table}[t]
\centering
\small
\caption{Headline results on all \numproblems\ problems, averaged over runs:
(a)~PBT mode, the leading metric; (b)~Baseline mode, a diagnostic.
\textbf{F$\geq 1$}: problems with at least one bug found; \textbf{Perfect}: problems
with all bugs found; \textbf{Turns}: tool calls per problem; $\boldsymbol{\Delta}$:
PBT$-$Baseline recall (pp). $\pm$ is the half-width of a $95\%$ percentile bootstrap
CI over (run, problem) instances ($1{,}000$ resamples). Best per column in bold.
$^{\dagger}$Claude Opus 4.8 was evaluated after review on the same OpenHands
scaffold with provider-default reasoning settings (PBT: 4 runs; Baseline: 2 runs;
$n=377$ PBT instances after excluding 23 trajectories stopped by a harness defect,
Appendix~\ref{app:post-review}).
$^{\ddagger}$Claude Code CLI driving Opus 4.8 with its default settings and the same PBT
prompt (3 runs, 15-minute limit); turns are not comparable across scaffolds.
To compare agents, use the paired problem-level test of Appendix~\ref{app:stability}.}
\label{tab:headline}
\setlength{\tabcolsep}{4pt}
\begin{tabular}{lrrrr}
\multicolumn{5}{l}{\textbf{(a) PBT mode} (Hypothesis-scaffolded prompt) --- leaderboard}\\
\toprule
\textbf{Model} & \textbf{Recall} & \textbf{F$\geq 1$} & \textbf{Perfect} & \textbf{Turns} \\
\midrule
Sonnet 4.6                   & \textbf{83.4\%}$\pm3.3$ & 92.7\%$\pm2.8$          & \textbf{67.0\%}$\pm5.2$ & 68.4 \\
Opus 4.8$^{\dagger}$         & 81.2\%$\pm2.8$          & 95.2\%$\pm2.1$          & 61.8\%$\pm5.0$          & 42.2 \\
\quad + Claude Code$^{\ddagger}$ & 78.6\%$\pm3.3$      & \textbf{95.7\%}$\pm2.2$ & 53.7\%$\pm5.8$          & --   \\
Qwen 3.6 Plus                & 78.0\%$\pm3.5$          & 90.3\%$\pm3.2$          & 56.0\%$\pm5.7$          & 45.1 \\
GLM 5.1                      & 72.1\%$\pm4.5$          & 77.0\%$\pm4.8$          & 63.0\%$\pm5.0$          & 89.2 \\
Gemini 3 Flash               & 62.8\%$\pm3.8$          & 84.0\%$\pm4.0$          & 37.0\%$\pm5.7$          & 34.0 \\
DeepSeek V3.2                & 61.0\%$\pm4.4$          & 79.7\%$\pm4.5$          & 38.7\%$\pm5.7$          & 50.9 \\
Step 3.5 Flash               & 58.1\%$\pm4.9$          & 70.7\%$\pm5.2$          & 41.0\%$\pm5.5$          & 83.4 \\
Qwen3.5-35B-A3B             & 54.3\%$\pm4.4$          & 77.3\%$\pm4.8$          & 25.7\%$\pm4.8$          & 48.1 \\
Grok 4.1 Fast                & 42.1\%$\pm4.3$          & 63.0\%$\pm5.5$          & 21.0\%$\pm4.7$          & 31.8 \\
\bottomrule
\end{tabular}

\vspace{6pt}
\begin{tabular}{lrrrrr}
\multicolumn{6}{l}{\textbf{(b) Baseline mode} (open-ended prompt) --- diagnostic}\\
\toprule
\textbf{Model} & \textbf{Recall} & \textbf{F$\geq 1$} & \textbf{Perfect} & $\boldsymbol{\Delta}$ & \textbf{Turns} \\
\midrule
Sonnet 4.6                   & \textbf{76.7\%}$\pm2.7$ & \textbf{98.0\%}$\pm1.5$ & 43.0\%$\pm5.8$          & $+6.7$  & 42.5 \\
GLM 5.1                      & 66.6\%$\pm4.3$          & 79.7\%$\pm4.5$          & \textbf{44.0\%}$\pm5.3$ & $+5.5$  & 57.4 \\
DeepSeek V3.2                & 64.2\%$\pm3.8$          & 90.0\%$\pm3.5$          & 35.3\%$\pm5.5$          & $-3.2$  & 61.8 \\
Gemini 3 Flash               & 56.2\%$\pm4.7$          & 71.3\%$\pm5.2$          & 38.3\%$\pm5.7$          & $+6.6$  & 44.6 \\
Qwen 3.6 Plus                & 53.5\%$\pm3.8$          & 85.3\%$\pm4.0$          & 18.3\%$\pm4.2$          & $+24.5$ & 27.1 \\
Grok 4.1 Fast                & 50.1\%$\pm3.8$          & 88.3\%$\pm3.5$          & 20.7\%$\pm4.7$          & $-8.0$  & 33.3 \\
Step 3.5 Flash               & 37.8\%$\pm3.7$          & 67.3\%$\pm5.3$          & 14.0\%$\pm4.0$          & $+20.3$ & 51.0 \\
Opus 4.8$^{\dagger}$         & 39.3\%$\pm3.6$          & 96.5\%$\pm2.5$          & 11.5\%$\pm4.5$          & $+41.9$ & 20.7 \\
Qwen3.5-35B-A3B             & 31.4\%$\pm2.9$          & 76.0\%$\pm4.7$          &  5.3\%$\pm2.7$          & $+22.9$ & 30.2 \\
\bottomrule
\end{tabular}
\end{table}

\paragraph{(1) A newer frontier model does not saturate the benchmark.}
The best PBT-mode recall is $83.4\%$ (Sonnet~4.6). Opus~4.8 reaches
$81.2\%\pm2.8$ on the same scaffold (paired difference to Sonnet~4.6: $-2.4$pp,
$95\%$ CI $[-7.2,+2.4]$); in Claude Code it reaches $78.6\%$ ($-2.4$pp $[-5.5,+0.7]$
vs.\ OpenHands; $-4.8$pp $[-9.6,-0.1]$ vs.\ Sonnet~4.6). No configuration finds all bugs on more than $67\%$ of
problems, and neither frontier configuration reliably finds the two bugs that no
original cell finds reliably (Section~\ref{subsec:ensemble}).

\paragraph{(2) The PBT$-$Baseline gap is a diagnostic, not a capability ranking.}
Among the eight original models, $\Delta$ is largest for models with weaker
baselines (Qwen~3.6 Plus $+24.5$pp, Qwen3.5-35B-A3B $+22.9$pp, Step~3.5 Flash
$+20.3$pp) and small or negative for the others (Sonnet~4.6 $+6.7$pp;
DeepSeek~V3.2 $-3.2$pp; Grok~4.1 Fast $-8.0$pp). A paired bootstrap over
problems gives $95\%$ CIs that exclude zero for the two Qwen models, Step,
Sonnet ($[+1.5,+11.4]$), and Grok ($[-15.0,-1.7]$), but not for GLM, Gemini, or
DeepSeek (Appendix~\ref{app:stability}). Opus~4.8 has the largest gap of all
($+41.9$pp): its Baseline recall is as low as the weakest models' while its PBT
recall matches the best. The reason is behavioural: under the open-ended Baseline prompt
(``write a test that fails''), Opus~4.8 stops after exposing one bug. It writes
a median of two test functions per problem (Sonnet~4.6: eight), finds at least
one bug on $97\%$ of problems, yet finds \emph{all} bugs on only $12\%$.
$\Delta$ therefore reflects how exhaustively a model answers an open-ended testing
request as much as its PBT skill, which is why PBT-mode recall is the leading
metric (Section~\ref{sec:experimental_setup}); Section~\ref{subsec:ablation}
decomposes what the PBT prompt contributes.

\paragraph{(3) Problem coverage and full recall diverge.}
Sonnet~4.6 Baseline has the highest problem coverage (98.0\%) but full recall of
only 43.0\%; PBT mode trades a small coverage drop (92.7\%) for a large gain in
full recall (67.0\%), consistent with its instruction to write multiple focused tests.

\subsection{Ensemble upper bound}
\label{subsec:ensemble}

To separate capability limits from problem diversity, we define \emph{reliable coverage}: a bug is reliably covered by a cell if found in at least two of three independent runs, smoothing over run-to-run variation.

Figure~\ref{fig:ensemble-curve} (Appendix~\ref{app:full-results}) shows the greedy union curve. The best
single cell is Claude Sonnet 4.6 PBT, which reliably covers 317 of 365
bugs ($86.8\%$); Qwen 3.6 Plus PBT follows at 303 ($83.0\%$). Greedily
adding a second cell (Claude Sonnet 4.6 Baseline) lifts union coverage
to $95.1\%$; the full 16-cell ensemble reaches $99.5\%$. Two bugs are never
reliably found by any cell, each only in single runs of a few cells
(Appendix~\ref{app:never-found}). They fail for different reasons: \texttt{BIDC-004}
bug~3 (L2) fires on most inputs once \texttt{update()} is called with a bidict
argument, a case agents rarely test (a coverage miss), whereas \texttt{TMLK-005}
bug~1 (L3) needs a TOML document that splits one table into three or more
interleaved sections (a search miss).
The frontier runs do not move this ceiling: Opus~4.8 PBT reliably covers
301 bugs in its first three runs ($82.5\%$), adding Opus~4.8 cells to the union contributes no new bug,
and \texttt{TMLK-005 bug\_1} is found in only one of three Opus~4.8 runs.
Within-model mode complementarity provides most of the ensemble headroom: the
same model in two modes (Sonnet PBT + Sonnet Baseline) accounts for 30 of the
46 marginal bugs (65\%); cross-architecture additions contribute modestly
(Step~+9, Gemini~+6, GLM~+1). PBT and Baseline thus elicit complementary
testing strategies even within a single architecture.

%% file: sections/analysis.tex
\section{Analysis}
\label{sec:analysis}

\subsection{What does the PBT prompt contribute? A controlled ablation}
\label{subsec:ablation}
\label{subsec:baseline-vs-pbt}

The Baseline and PBT regimes differ in several ways at once: the nominal
per-test timeout (120\,s vs.\ 300\,s), whether Hypothesis is preinstalled and required,
the task goal (``write a failing test'' vs.\ ``cover the documented
specification''), and the PBT prompt's property taxonomy, ``5+ functions'' target,
and step-by-step template.

\paragraph{The per-test timeout has no effect.}
In both regimes the agent SDK's command interface returns after 30\,s, so the
nominal 120\,s and 300\,s pytest timeouts never bind and both regimes share an
effective 30\,s limit per test function. Re-scoring all 2,092 released Baseline
instances with a test file under the PBT setting confirms this: only 2 of 7,699
per-bug verdicts changed, both in one instance and due to test nondeterminism.

\paragraph{Framework versus taxonomy and template.}
The framework-only prompt keeps the PBT prompt's framework instructions (Hypothesis
preinstalled, \texttt{@given} on every test, tests derived from the documentation,
``cover the specification rather than hunt for one bug'') and drops the taxonomy,
target, and template, separating (i)~the framework instructions as a bundle from
(ii)~the taxonomy and template (Table~\ref{tab:ablation}).
For Opus~4.8, whose three conditions share one infrastructure, the framework
instructions account for essentially the whole gain ($+41.5$pp vs.\ $+0.2$pp);
since its Baseline failure is stopping after one bug, the coverage goal is the
likely active ingredient, which the bundle cannot separate from Hypothesis itself.
For Qwen~3.6 Plus the framework instructions contribute more than the taxonomy and
template ($+17.8$ vs.\ $+6.7$pp); two paraphrases of the framework-only prompt give
$76.1\%$ and $72.4\%$ vs.\ $71.3\%$ (Appendix~\ref{app:post-review}). For
DeepSeek~V3.2 ($22\%$ of framework-only trajectories lack a test file, mostly from
malformed tool calls) the taxonomy and template add $+11.3$pp ($+9.1$pp on problems
with a file); for Sonnet~4.6 neither difference is significant.
These three models' framework-only cells ran months later, on different machines,
with fewer runs; we treat their rows as indicative. In short, Opus~4.8 does not need to be taught \emph{how}
to write property tests but does not test the specification exhaustively unless
asked, while mid-capability models also gain from being told \emph{which}
properties to look for.

\begin{table}[t]
\centering
\small
\caption{Controlled prompt ablation. Mean bug recall (\%) on problems evaluated
under all three conditions, and paired differences in pp with $95\%$ bootstrap
CIs over problems ($2{,}000$ resamples). Runs are Baseline/framework-only/full-PBT.
\emph{Top:} all common problems (a missing test file counts as zero recall);
\emph{bottom:} problems where every run of every condition produced a test file.
No-file is the share of framework-only trajectories without a test file.
Opus~4.8 pools all its OpenHands runs.}
\label{tab:ablation}
\setlength{\tabcolsep}{3.5pt}
\begin{tabular}{lcrrrrrrr}
\toprule
\textbf{Model} & \textbf{Runs} & $n$ & \textbf{Base.} & \textbf{Frame.} & \textbf{PBT} & \textbf{Frame.$-$Base.} & \textbf{PBT$-$Frame.} & \textbf{No-file} \\
\midrule
Opus 4.8      & 2/3/4 & 100 & 39.3 & 80.8 & 81.0 & $+41.5$ {\scriptsize[$+34.2,+47.7$]} & $+0.2$ {\scriptsize[$-3.4,+3.7$]}   & 0.4\% \\
Qwen 3.6 Plus & 3/3/3 & 100 & 53.5 & 71.3 & 78.0 & $+17.8$ {\scriptsize[$+11.4,+24.1$]} & $+6.7$ {\scriptsize[$+2.2,+11.6$]}  & 3.9\% \\
Sonnet 4.6    & 3/1/3 & 78  & 78.0 & 79.3 & 84.9 & $+1.3$ {\scriptsize[$-7.2,+8.7$]}    & $+5.6$ {\scriptsize[$-0.9,+12.9$]}  & 7.7\% \\
DeepSeek V3.2 & 3/2/3 & 99  & 64.1 & 49.6 & 60.9 & $-14.4$ {\scriptsize[$-22.9,-6.3$]}  & $+11.3$ {\scriptsize[$+4.9,+18.0$]} & 21.6\% \\
\midrule
\multicolumn{9}{l}{\emph{Problems with a test file in all runs of all conditions}} \\
Opus 4.8      & 2/3/4 & 93  & 40.4 & 80.7 & 82.0 & $+40.2$ {\scriptsize[$+33.6,+46.6$]} & $+1.4$ {\scriptsize[$-2.1,+4.7$]}  & -- \\
Qwen 3.6 Plus & 3/3/3 & 56  & 62.0 & 80.4 & 86.4 & $+18.4$ {\scriptsize[$+11.4,+25.5$]} & $+6.0$ {\scriptsize[$+1.6,+10.6$]}  & -- \\
Sonnet 4.6    & 3/1/3 & 65  & 80.8 & 85.1 & 88.9 & $+4.4$ {\scriptsize[$-3.5,+12.1$]}   & $+3.8$ {\scriptsize[$-1.4,+9.2$]}   & -- \\
DeepSeek V3.2 & 3/2/3 & 42  & 59.3 & 62.2 & 71.3 & $+2.9$ {\scriptsize[$-9.8,+15.4$]}   & $+9.1$ {\scriptsize[$+1.0,+18.0$]}  & -- \\
\bottomrule
\end{tabular}
\end{table}

\paragraph{Implications for model development.}
Training signals that reward exposing \emph{every} specification violation (full
recall) rather than one failing test target Opus~4.8's gap. Most sampled PBT-mode
failures (62\%) are search failures (below); the construction pipeline of
Section~\ref{sec:benchmark} can generate injected bugs with F$\to$P labels on
held-out libraries to supervise this skill, keeping \benchname\ itself out of training data.

\subsection{Diagnostics: difficulty, complementarity, failure modes}
\label{subsec:diagnostics}

Three short diagnostics support the headline numbers (full tables in
Appendices~\ref{app:difficulty-empirical}--\ref{app:failures-table}).
\emph{(1)~L1--L3 labels track empirical hardness:} averaged across the eight
original models, recall is $66\%/60\%/51\%$ on L1/L2/L3, and the ordering holds in
15 of 16 cells (Table~\ref{tab:recall-by-difficulty}) and for Opus~4.8
($88\%/83\%/76\%$).
The L1/L2 gap is narrower than the L2/L3 gap: L1 and L2 both require reasoning
about input constraints, while L3 additionally requires understanding protocols
that span several API calls.
\emph{(2)~No model dominates the hardest bugs:} the three \texttt{GALS}
finite-field problems are solved at different rates by different model families,
so a cross-model union is required to cover them.
\emph{(3)~Baseline and PBT fail differently.} In 20 sampled failures per cell,
Baseline failures are mostly \textbf{incorrect assertions} (59\%; the test's
expected value is itself wrong) and \textbf{overly concrete tests} (32\%; a few
hand-picked inputs that miss the trigger region). PBT failures are mostly search
failures: \textbf{assume misuse} (31\%, up to 75\% for GLM~5.1), where the agent
uses \texttt{hypothesis.assume()}, which discards generated inputs that fail a
predicate, to exclude exactly the inputs that trigger the bug; and \textbf{wrong
strategy range} (31\%, up to 50\% for Gemini~3 Flash), where the generator's
value range does not cover the trigger region. The per-difficulty pattern of
$\Delta$ is analyzed in Appendix~\ref{app:difficulty-empirical}.

\paragraph{Integrity and stability.}
\label{subsec:adversarial}
\label{subsec:sensitivity}
Under an explicit cheat prompt adapted from Terminal-Bench's audit
protocol~\citep{merrill2026terminalbench}, 30 Sonnet~4.6 trajectories reached
normal recall ($0.85$--$0.93$) with zero attack signatures such as hard-coded
assertions or version checks. An early harness
version left \texttt{*.orig} patch backups readable; all affected trajectories
($8.0\%$) were re-run, with no detectable effect on recall (Wilcoxon $p=0.31$;
Appendix~\ref{app:leak-method}). The median CI half-width across the 16 original
cells is $3.8$\,pp; run-to-run variance comes from agents writing different
tests and, in PBT mode, from Hypothesis's random search (Appendix~\ref{app:stability}).

%% file: sections/conclusion.tex
\section{Conclusion}
\label{sec:conclusion}

\benchname\ probes a capability existing code-agent benchmarks
implicitly assume: moving from a documented invariant to a Hypothesis
strategy precise enough to expose a semantic bug. On \numproblems\
curated problems across \numlibraries\ real Python libraries,
property-based testing remains a persistent capability gap: the best
agent reaches $83\%$ bug recall, a newer frontier model and a commercial agent
scaffold do not improve on it, and two of $365$ bugs are never reliably found.
How much an agent gains from being asked to use PBT is not monotone in model
strength; for Opus~4.8 the gain comes from the framework instructions rather than
the property taxonomy. We recommend PBT-mode recall as the leading metric, with
full recall and the Baseline mode as diagnostics.
\benchname\ complements test-synthesis benchmarks mined from pull requests
(\textsc{TestExplora}, \textsc{SWT-Bench}) and in-the-wild agentic
PBT~\citep{agenticpbt2025} by measuring, against ground truth, exactly the bug
distribution for which PBT is the intended tool.

\emph{Limitations.}
Curated bugs trade population fidelity for PBT-testability
(\S\ref{subsec:fidelity}), and \numproblems\ problems in one language limit
coverage of real software diversity; \benchname\ covers only Python libraries
with English documentation. The construction pipeline is language-agnostic, and
we outline its extension to more libraries and to languages with mature PBT
frameworks in Appendix~\ref{app:scaling}.
Most agents run on one scaffold (OpenHands); the frontier and Claude Code
results come from one model vendor, and we could not evaluate Codex CLI. The ablation's framework arm bundles Hypothesis
with a coverage-oriented goal, and three of its models ran later on different
infrastructure. Prompt sensitivity is only partially characterized
(Appendix~\ref{app:stability}), and the \texttt{max\_examples=200} budget likely
caps L2/L3 recall on narrow-trigger problems (e.g.\ \texttt{GALS-003}).
\emph{Future work:} a \texttt{max\_examples} sweep, and prompts that warn against \texttt{assume()}-based filtering.

%% file: sections/acknowledgments.tex
%

SSD acknowledges the support of NSF IIS 2229881, NSF CIF 2212261, NSF IIS 2143493, UW-Tsukuba Amazon NVIDIA Cross Pacific AI Initiative (XPAI), Google TPU Builders Program, and the AI2050 program at Schmidt Sciences.

\textbf{Competing interests.} The authors declare no competing interests.

%% file: sections/appendix.tex
\section{Additional experimental details}
\label{app:details}

\subsection{Problem format}
\label{app:problem-format}

Each problem is stored in \texttt{libraries/<lib>/problems/<ID>/} and consists of five components.
\texttt{problem.yaml} defines metadata including library version, injected bugs with difficulty levels (L1--L3), trigger conditions, ground-truth Hypothesis strategies, and semantic tags (\texttt{prop:*}, \texttt{bug:*}, \texttt{strat:*}).
Each injected bug is represented as a \texttt{bug\_N.patch} unified diff applied at container startup, enabling reversible injection: the default state includes all patches; removing patch $N$ yields the fixed-bug-$N$ configuration for per-bug scoring.
\texttt{docs/} provides agent-accessible documentation (official sources when available), serving as the only external oracle.
\texttt{existing\_tests/} contains unit tests that pass on the buggy version, ensuring injected bugs escape the baseline suite.
\texttt{ground\_truth/} includes reference solutions and strategy specifications used exclusively for offline validation and never exposed to the agent.

\subsection{Hypothesis strategy primer}
\label{app:hypothesis-primer}

In Hypothesis, a \emph{strategy} (\texttt{SearchStrategy}) is an object that draws random values from a defined search space and automatically shrinks any discovered counterexample to a minimal failing form.
The \texttt{hypothesis.strategies} module (aliased \texttt{st} by convention) provides built-in primitives --- \texttt{st.lists()}, \texttt{st.integers()}, \texttt{st.text()}, \texttt{st.floats()} --- and the \texttt{@st.composite} decorator for constructing structured, dependent strategies.
The \texttt{@given(\textit{strat},~\ldots)} decorator binds one or more strategies to a test function's arguments and drives the adaptive search engine.
An example using \texttt{st.lists()} is shown in Figure~\ref{fig:pbt-qsort-bug}.

\subsection{Prompt templates}
\label{app:prompts}

Below we present abridged versions of the prompt templates used in the evaluation (full templates with Jinja2 conditionals are in \texttt{eval/prompts/} in the released repository). Per-problem variables (library name, version, scope files) are filled at runtime.

\subsubsection*{Baseline prompt}

\begin{minted}[fontsize=\scriptsize, breaklines]{text}
You have been given a Python library to test. The library implementation
may have one or more issues that the existing test suite does not detect.

## Your task

Write a test that **fails** when run against the current (potentially buggy)
version of the library, and would **pass** after the issue is fixed.
Save your final test to `/workspace/pbt_test.py`.

## Workspace layout

/workspace/
  lib/                 -- the library under test (may contain a bug)
  docs/                -- API documentation and usage guide
  existing_tests/      -- existing test suite (all currently pass; do NOT modify)
  pytest.ini           -- pre-configured: lib/ is already on the Python path
  pbt_test.py          -- write your test here

## Instructions

Phase 1 -- Understand the library
1.0 Inspect the scope files listed under Scope.
1.1 Read the documentation in /workspace/docs/.
1.2 Study the existing tests to understand what is already covered.
1.3 Identify the public API: what functions/methods exist, what are their contracts?

Phase 2 -- Find uncovered behavior
2.1 Think about what behaviors the existing tests do NOT cover.
2.2 Consider boundary conditions, edge cases, and properties that should always hold.
2.3 Look for semantic invariants: properties that should be true for all valid inputs.

Phase 3 -- Write a failing test
3.1 Write a test using pytest that exercises uncovered behavior.
3.2 The test should fail because of the bug in the current library.
3.3 Run: timeout 60 python -m pytest /workspace/pbt_test.py --tb=short -v
3.4 If the test passes, rethink your approach and return to Phase 2.

Phase 4 -- Finalize
4.1 Make sure the test file is saved at /workspace/pbt_test.py.
4.2 Confirm the test fails with a meaningful assertion error.
4.3 Do NOT modify any files in /workspace/lib/ or /workspace/existing_tests/.

Notes:
- Issues are subtle: the library mostly works, but fails silently on certain inputs.
- There may be more than one issue. Try to cover as many as possible.
- Write focused test functions: one specific potential issue per function.
\end{minted}

\subsubsection*{PBT (Hypothesis-scaffolded) prompt}

\begin{minted}[fontsize=\scriptsize, breaklines]{text}
You have been given a Python library to test using Property-Based Testing (PBT)
with the Hypothesis framework. The library implementation may have one or more
issues that the existing test suite does not detect.

Your goal is NOT to find the bug first and then write a test for it. Instead,
write the most comprehensive set of property-based tests you can derive purely
from the documentation -- covering every behavior, invariant, and edge case
described. If the implementation deviates from its specification anywhere,
Hypothesis will find it.

## Strategy

Step 1 -- Read the documentation thoroughly
Read all files in /workspace/docs/. For each function or class in scope, extract:
- The precise semantics: what does it do, what does it return?
- All stated invariants: what should always be true?
- Edge cases mentioned explicitly
- Relationships between functions: roundtrip properties, monotonicity
- Any numeric thresholds, size limits, or type-specific behavior

Do NOT read the source code at this stage. The documentation is your oracle.

Step 2 -- Write as many property tests as possible
For every documented behavior or invariant, write a @given test. Aim for 5+
distinct test functions. Prefer:
- Roundtrip properties: decode(encode(x)) == x
- Model-based oracles: compare to a simple Python reference
- Invariant preservation: after any operations, a structural invariant holds
- Commutativity / associativity / idempotence
- Monotonicity: larger input -> larger (or smaller) output
- Boundary behavior: empty collections, single elements, min/max values

Use @settings(max_examples=200, deadline=None) on every test.

Step 3 -- Cover edge cases explicitly
Beyond general properties, write targeted tests for documented edge cases.

Step 4 -- Run and save
Run: python -m pytest /workspace/pbt_test.py --tb=short -v
Tests that fail immediately indicate you found a bug -- excellent.
Tests that pass are still valuable. Save all tests regardless.

## Rules
- Do NOT modify /workspace/lib/ or /workspace/existing_tests/.
- Every test must use @given from Hypothesis (not just fixed inputs).
- Write multiple focused functions -- one property per function.
- The documentation is your primary oracle.
- Do not spend time hunting for a specific bug. Write tests that cover
  the spec broadly; Hypothesis will find the deviations.
\end{minted}

\subsubsection*{Framework-only prompt (post-review ablation)}

Identical environment to the PBT prompt; the property taxonomy, the ``5+
functions'' target, and the step-by-step template are removed.

\begin{minted}[fontsize=\scriptsize, breaklines]{text}
You have been given a Python library to test using Property-Based Testing (PBT)
with the Hypothesis framework. The library implementation may have one or more
issues that the existing test suite does not detect.

Your goal is not to find the bug first and then write a test for it. Instead,
write property-based tests derived purely from the documentation. If the
implementation deviates from its specification anywhere, Hypothesis will find it.

[workspace layout, library, scope, and setup notes as in the PBT prompt]

## Task
Read the documentation in /workspace/docs/ and write property-based tests for
the library, using the Hypothesis framework.
- Every test must use the @given decorator from Hypothesis (not fixed-input tests).
- Use @settings(max_examples=200, deadline=None) on every test.
- The documentation is your primary oracle. Do not assume the source code is
  correct, and do not read /workspace/lib/ while deriving your tests.
- Do NOT modify /workspace/lib/ or /workspace/existing_tests/.
- Do not spend time hunting for a specific bug. Write tests that cover the
  specification; Hypothesis will find the deviations.

Run your tests and save them to /workspace/pbt_test.py:
  python -m pytest /workspace/pbt_test.py --tb=short -v
Save all tests regardless of whether they currently fail or pass.
\end{minted}

\subsection{Library list}
\label{app:library-list}

The \numlibraries\ canonical libraries span seven thematic domains:
\emph{serialization} (msgpack, cbor2, construct, yaml, tomlkit, lark,
pycparser, pypdf); \emph{data structures} (sortedcontainers, bidict,
bintrees, cachetools, diskcache, intervaltree, multiset, boltons, portion, more-itertools, toolz); \emph{date--time}
(arrow, dateutil, babel, icalendar, pendulum); \emph{type systems and schemas}
(attrs, jsonschema, cattrs, marshmallow, pyrsistent); \emph{numerics}
(mpmath, pint, galois, sympy, networkx); \emph{state machines} (transitions);
and \emph{parsing/HTML} (pyparsing, parso, html5lib, openpyxl, pyasn1).
Each problem's \texttt{problem.yaml} pins the library version; per-bug metadata
(library, difficulty, description, tags) for the 365 canonical bugs and for
retired problems is in \texttt{paper/analysis/bug\_metadata.csv}.

\subsection{Bug-class to real-world analogue mapping}
\label{app:bug-class-map}

Table~\ref{tab:bug-class} groups the 365 injected bugs into eleven
high-level classes derived from per-bug \texttt{tags} entries in
\texttt{problem.yaml}. Each class corresponds to a bug pattern
commonly observed in open-source Python projects.

\begin{table}[h]
\centering
\small
\caption{Injected-bug taxonomy across the \numproblems\ canonical
problems. Columns are per-bug difficulty L1--L3.}
\label{tab:bug-class}
\begin{tabular}{lrrrr}
\toprule
Class & L1 & L2 & L3 & Total \\
\midrule
Off-by-one / boundary        & 23 & 30 & 12 & 65 (18\%) \\
Wrong reference / expression & 12 & 25 & 19 & 56 (15\%) \\
Condition / logic inversion  & 12 & 30 &  7 & 49 (13\%) \\
Wrong operator               & 13 & 28 &  3 & 44 (12\%) \\
Argument / field order swap  &  6 & 19 & 17 & 42 (12\%) \\
Missing operation            &  2 & 19 &  8 & 29 ( 8\%) \\
Traversal / ordering         &  4 & 10 &  5 & 19 ( 5\%) \\
Sign / direction error       &  4 &  6 &  6 & 16 ( 4\%) \\
Protocol / domain-specific   &  0 &  6 &  9 & 15 ( 4\%) \\
Wrong constant / value       &  7 &  5 &  2 & 14 ( 4\%) \\
State corruption / mutation  &  2 &  2 &  4 &  8 ( 2\%) \\
Other                        &  2 &  4 &  2 &  8 ( 2\%) \\
\midrule
Total                        &  87 & 184 & 94 & 365 \\
\bottomrule
\end{tabular}
\end{table}

\subsection{Complete per-cell results heatmap}
\label{app:full-results}

\begin{figure}[h]
\centering
\includegraphics[width=0.85\linewidth]{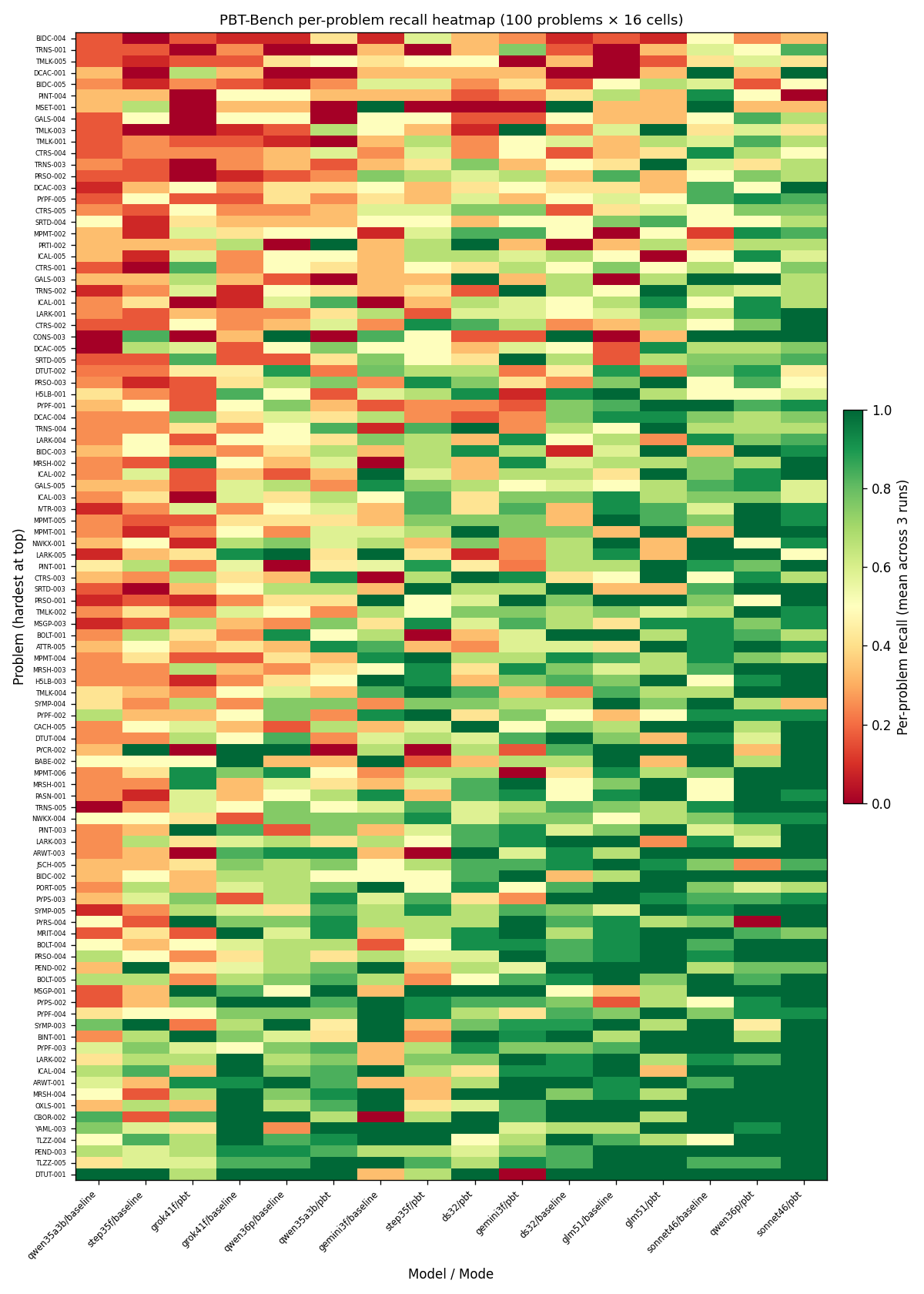}
\caption{Per-problem recall, 100 problems $\times$ 16 cells. Problems
sorted by mean cross-cell recall (hardest at top). Cells sorted by
overall recall (weakest at left). Color is per-problem recall averaged
across 3 runs. Column labels use the short model names of the released
\texttt{results.csv} (\texttt{qwen35a3b} denotes Qwen3.5-35B-A3B).}
\label{fig:per-problem-heatmap}
\end{figure}

Figure~\ref{fig:per-problem-heatmap} shows per-problem recall across
all \numproblems\ problems (rows, sorted hardest at top) and all
sixteen (model, mode) cells (columns, sorted weakest at left). Each
cell color is the mean recall across the three runs for that cell,
using the merged dataset (iter2 with clean-rerun substitutions where
available). The ten hardest problems by mean cross-cell recall
(\texttt{BIDC-004} $\to$ \texttt{TMLK-001}, with \texttt{BIDC-004}
scoring $22.4\%$) dominate the ensemble-upper-bound ceiling in
Section~\ref{subsec:ensemble}. The full underlying CSV is released at
\texttt{paper/analysis/hardest\_problems.csv}.

\begin{figure}[h]
\centering
\includegraphics[width=0.7\linewidth]{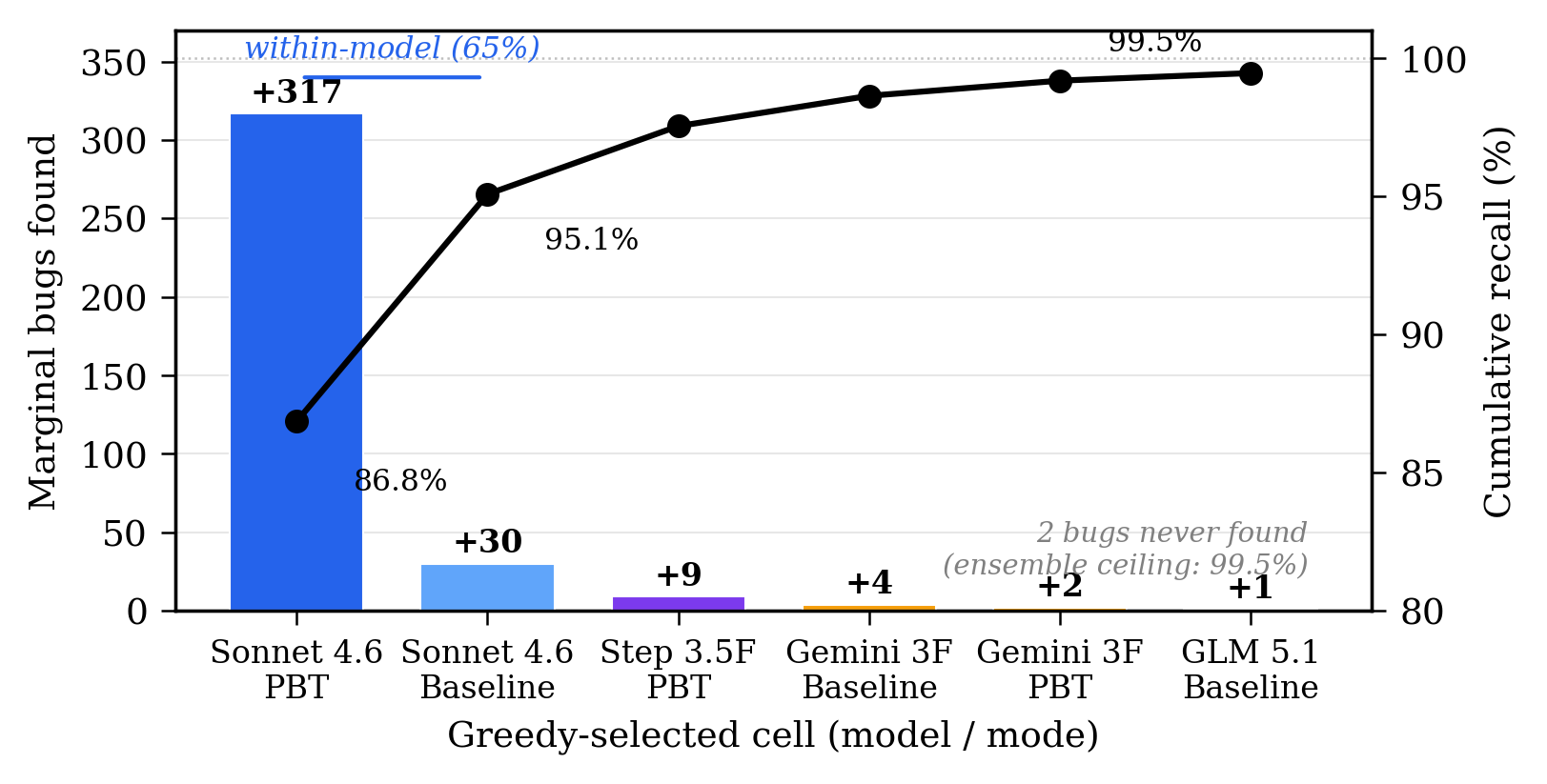}
\caption{Greedy ensemble construction over the sixteen original cells:
  marginal bugs found per cell (bars, left axis) and cumulative recall (line,
  right axis). Only 6 of 16 cells contribute unique bugs; within-model mode
  complementarity (Sonnet PBT + Baseline) accounts for 65\% of the marginal gains.}
\label{fig:ensemble-curve}
\end{figure}

\subsection{The \texttt{.orig} leak: detection and rerun policy}
\label{app:leak-method}

The initial harness (\texttt{patch -p1}) occasionally left \texttt{*.orig} backup files readable by the agent. We fixed this in commit \texttt{1048fd6} by unconditionally deleting backups after patch application.

\paragraph{Detection.} We flag a workspace as ``exploited'' if its \texttt{chat.md} contains any regex match against five read-action patterns targeting \texttt{.orig} files (shell \texttt{diff}, \texttt{cat}/\texttt{head}/\texttt{tail}, \texttt{grep}, Python \texttt{open}, or \texttt{file\_editor view}). In total, $383$ of $4{,}800$ workspaces ($8.0\%$) matched.

\paragraph{Rerun policy.} If \emph{any} of the three runs for a (model, mode, problem) tuple contained exploitation, we reran \emph{all three} runs on the post-fix harness. This affected $224$ tuples ($672$ workspaces) across six models. The merged dataset used for all reported numbers contains zero residual pre-fix rows for any contaminated tuple.

\paragraph{Statistical effect.} Within-problem paired tests (Wilcoxon signed-rank $p=0.31$, $n=160$ mixed tuples) show no detectable causal effect of exploitation on recall. The rerun is a hygiene measure, not a correction for inflated scores.

\subsection{Failure-mode taxonomy: method and examples}
\label{app:failures}

\paragraph{Classifier.}
Failures were classified by a deterministic rule cascade implemented in \texttt{paper/analysis/build\_failure\_taxonomy.py}. The classifier consumes three signals per failed trial: (i)~the submitted \texttt{pbt\_test.py} source, examined for \texttt{@given}, \texttt{assume(\ldots)} calls, default-strategy constructors like \texttt{st.integers()} or \texttt{st.text()}, and weak-assertion patterns (\texttt{is not None}, \texttt{len(x) >= 0}, \texttt{isinstance}, round-trip
encode/decode); (ii)~the per-function pass pattern on
\texttt{(buggy, fixed)} library versions from
\texttt{bugs\_result.json}; (iii)~\texttt{chat.md} for the complete agent interaction history. The cascade and its outputs across 320 sampled failures are in \texttt{paper/analysis/failure\_taxonomy.csv}, where Incorrect Assertion is labelled \texttt{WRONG\_API\_USE}.

\paragraph{Sampling.}
For each of the 16 (model, mode) cells we sampled 20 failures uniformly
over problems, subject to \texttt{test\_file\_found=True AND found=False}.
Across cells: 320 samples. The sample was drawn from the trial table before the
\texttt{.orig}-leak reruns (Appendix~\ref{app:leak-method}) were merged, so 28
sampled trajectories belong to tuples that were later rerun; the released
trajectories for these tuples are the reruns, in which 11 of the sampled bugs are
found. Dropping the 28 samples leaves every cell's dominant category unchanged
and moves the pooled per-mode shares by at most 1\,pp.

\paragraph{Representative examples.}
\begin{itemize}[leftmargin=*, topsep=2pt, itemsep=2pt]
\item \textbf{Assume Misuse} --- Sonnet 4.6 PBT on \texttt{H5LB-003}
  bug 2: \texttt{assume(val1 != val2)} filters out the exact
  duplicate-value case that is the bug's trigger.
\item \textbf{Assume Misuse} --- Sonnet 4.6 PBT on \texttt{SRTD-005}
  bug 2: \texttt{assume(query not in sl)} --- the bug is specifically
  about duplicate-query lookups.
\item \textbf{Wrong Strategy Range} --- DeepSeek V3.2 PBT on
  \texttt{DCAC-005}: \texttt{@given(st.integers())} for a cache-expire
  timestamp, making triggers at boundary seconds astronomically
  unlikely.
\item \textbf{Wrong Strategy Range} --- DeepSeek V3.2 PBT on
  \texttt{CTRS-001}: \texttt{st.text(alphabet=}\allowbreak\texttt{st.characters(whitelist\_categories=}\allowbreak\texttt{('L','N','P','S','Z')))} excludes control characters
  that the bug targets.
\item \textbf{Flaky Oracle} --- Grok 4.1 Fast Baseline on
  \texttt{ICAL-003}: oracle reads the serialized string back via
  \texttt{Calendar.from\_ical(ical\_str.encode())} --- the same buggy
  parser that produced the value.
\item \textbf{Incorrect Assertion} --- DeepSeek V3.2 Baseline on
  \texttt{H5LB-003}: a test asserts that \texttt{\&\#x41;} parses to
  \texttt{'A'}, but the test's expected value is itself what the buggy
  \texttt{consumeNumberEntity} produces --- fails on both buggy and
  fixed libraries.
\item \textbf{Setup Error} --- DeepSeek V3.2 PBT on \texttt{SYMP-005}:
  duplicate \texttt{from hypothesis import strategies as st} and
  \texttt{import hypothesis.strategies as st} produce a
  collection-time crash; test functions never run.
\item \textbf{Under-Specified Property} --- DeepSeek V3.2 Baseline on
  \texttt{PYCR-002}: \texttt{assert return\_match is not None} checks
  only the existence of some return statement, silent on whether it
  has the right shape.
\end{itemize}

\paragraph{Model-specific observations.}
\begin{itemize}[leftmargin=*, topsep=2pt, itemsep=1pt]
\item \textbf{Sonnet 4.6} and \textbf{GLM 5.1} over-use \texttt{assume()}: they understand Hypothesis well enough to write generators, but also willingly filter out the trigger region.
\item \textbf{Qwen3.5-35B-A3B} and \textbf{Gemini 3 Flash} mostly fail on
  unbounded default strategies (45--50\% Wrong Strategy Range).
\item \textbf{Step 3.5 Flash Baseline} skews Overly Concrete (80\%) ---
  it writes a handful of hardcoded examples when it could write a
  property test.
\end{itemize}

\paragraph{Taxonomy adequacy.}
After two rule-cascade refinements (splitting baseline ``overly
concrete'' failures by their \texttt{(buggy, fixed)} pass pattern; excluding \texttt{from hypothesis import assume} false-positives from the Assume-Misuse regex), zero failures landed in the OTHER bucket. We take this as evidence that the eight categories are collectively sufficient for PBT failure analysis at the granularity of a deterministic classifier.

\subsection{L1--L3 difficulty validation (empirical)}
\label{app:difficulty-empirical}

\begin{figure}[h]
\centering
\includegraphics[width=0.45\linewidth]{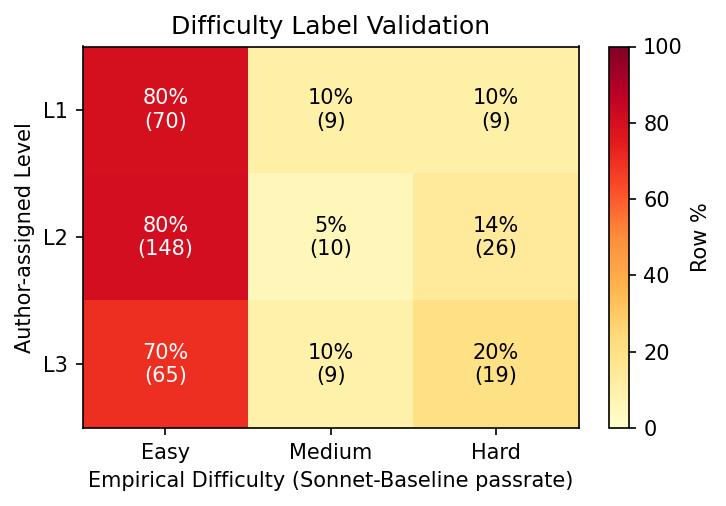}
\caption{Author-assigned difficulty (rows) against empirical difficulty
bucketed by Sonnet-Baseline passrate (columns). Each row sums to 100\%.
Cell labels are \emph{\%}$\pm$(count). L3 bugs concentrate in the Hard
bucket; L1 bugs concentrate in Easy.}
\label{fig:difficulty-matrix}
\end{figure}

Figure~\ref{fig:difficulty-matrix} plots the confusion matrix of
author-assigned L1--L3 against empirical difficulty buckets (Easy:
Sonnet-Baseline passrate $\geq 2/3$; Medium: $\geq 1/3$; Hard: $< 1/3$).
Mean Sonnet-Baseline passrate is $79\%$ for L1 bugs ($n=87$), $78\%$
for L2 ($n=184$), and $68\%$ for L3 ($n=94$). Medians are $1.0$ for all three levels
(Sonnet Baseline finds these on all three runs when it finds them at all),
so the difficulty signal is carried by the mean, not the median; the
distinction emerges in the left tail of each distribution.

The L1/L2 passrate gap is tighter ($79\%$ vs.\ $78\%$) than the L2/L3 gap ($78\%$ vs.\ $68\%$). This is expected: L1 and L2 both involve input-constraint reasoning (differing in the number of simultaneous constraints), while L3 additionally requires cross-function protocol understanding --- a qualitatively different skill. The gap widens when averaging across all models (L1: $66\%$, L2: $60\%$, L3: $51\%$), confirming that the three levels separate empirically distinct difficulty strata. Labels are assigned prospectively at design time, before any model evaluation, to avoid circularity.

Table~\ref{tab:recall-by-difficulty} reports per-cell recall broken down by
difficulty level. In 15 of 16 cells, recall decreases monotonically with
difficulty (L1 $>$ L2 $>$ L3); the sole exception is GLM~5.1 Baseline, where
L2 exceeds L1 by 2.2\,pp --- within the bootstrap CI width.

\begin{table}[h]
\centering
\small
\caption{Bug recall (\%) by difficulty level for all 16 (model, mode) cells,
sorted by PBT overall recall (descending). $\Delta$ = PBT$-$Baseline in pp;
negative values indicate PBT hurts. $n$: L1\,=\,87, L2\,=\,184, L3\,=\,94 bugs.
\emph{Note:} recall here is bugs-found / bugs-total pooled across all bugs in each difficulty stratum (bug-level average), whereas Table~\ref{tab:headline} reports problem-level average recall (each problem weighted equally); the two metrics differ slightly when problems contain unequal numbers of bugs.}
\label{tab:recall-by-difficulty}
\setlength{\tabcolsep}{5pt}
\begin{tabular}{llrrrr}
\toprule
\textbf{Model} & \textbf{Mode} & \textbf{L1} & \textbf{L2} & \textbf{L3} & \textbf{Overall} \\
\midrule
Sonnet 4.6     & PBT           & 93.1 & 83.2 & 74.5 & 83.3 \\
               & Baseline      & 79.3 & 77.5 & 68.1 & 75.5 \\
               & \textit{$\Delta$} & \textit{$+13.8$} & \textit{$+5.7$} & \textit{$+6.4$} & \textit{$+7.8$} \\
\midrule
Qwen 3.6 Plus  & PBT           & 87.0 & 79.3 & 67.7 & 78.2 \\
               & Baseline      & 61.7 & 53.8 & 45.0 & 53.4 \\
               & \textit{$\Delta$} & \textit{$+25.3$} & \textit{$+25.5$} & \textit{$+22.7$} & \textit{$+24.8$} \\
\midrule
GLM 5.1        & PBT           & 75.9 & 74.5 & 68.4 & 73.2 \\
               & Baseline      & 68.6 & 70.8$^*$ & 62.1 & 68.0 \\
               & \textit{$\Delta$} & \textit{$+7.3$} & \textit{$+3.7$} & \textit{$+6.3$} & \textit{$+5.2$} \\
\midrule
Gemini 3 Flash & PBT           & 70.9 & 67.8 & 53.9 & 64.9 \\
               & Baseline      & 62.1 & 54.9 & 54.3 & 56.4 \\
               & \textit{$\Delta$} & \textit{$+8.8$} & \textit{$+12.9$} & \textit{$-0.4$} & \textit{$+8.5$} \\
\midrule
DeepSeek V3.2  & PBT           & 70.5 & 61.8 & 49.6 & 60.7 \\
               & Baseline      & 67.8 & 65.8 & 57.4 & 64.1 \\
               & \textit{$\Delta$} & \textit{$+2.7$} & \textit{$-4.0$} & \textit{$-7.8$} & \textit{$-3.4$} \\
\midrule
Step 3.5 Flash & PBT           & 65.9 & 59.4 & 52.1 & 59.1 \\
               & Baseline      & 43.3 & 37.1 & 29.4 & 36.6 \\
               & \textit{$\Delta$} & \textit{$+22.6$} & \textit{$+22.3$} & \textit{$+22.7$} & \textit{$+22.5$} \\
\midrule
Qwen3.5-35B-A3B   & PBT           & 69.3 & 56.2 & 41.8 & 55.6 \\
               & Baseline      & 34.1 & 30.8 & 27.0 & 30.6 \\
               & \textit{$\Delta$} & \textit{$+35.2$} & \textit{$+25.4$} & \textit{$+14.8$} & \textit{$+25.0$} \\
\midrule
Grok 4.1 Fast  & PBT           & 52.5 & 43.7 & 29.8 & 42.2 \\
               & Baseline      & 58.2 & 48.9 & 40.4 & 48.9 \\
               & \textit{$\Delta$} & \textit{$-5.7$} & \textit{$-5.2$} & \textit{$-10.6$} & \textit{$-6.7$} \\
\bottomrule
\multicolumn{6}{l}{\footnotesize $^*$Sole non-monotone cell (L2 $>$ L1 by 2.2\,pp; within bootstrap CI).}
\end{tabular}
\end{table}

\paragraph{Per-difficulty $\Delta$ patterns.}

The following observations are exploratory: per-difficulty $\Delta$ values are not accompanied by per-cell confidence intervals, so small differences should be interpreted cautiously.

\textbf{(a) Diminishing-return pattern.}
Qwen3.5-35B-A3B shows the clearest gradient: $+35.2$/$+25.4$/$+14.8$\,pp on L1/L2/L3.
One interpretation is that the scaffold supplies a framework choice and input-constraint template sufficient for L1 bugs (single-constraint boundary checks) but insufficient for L3 bugs requiring cross-function protocol reasoning.
This pattern is specific to Qwen3.5-35B-A3B; Qwen~3.6~Plus ($+25.3$/$+25.5$/$+22.7$\,pp) is near-flat and grouped with the uniform-benefit pattern below.

\textbf{(b) Uniform-benefit pattern.}
Step~3.5~Flash and Qwen~3.6~Plus achieve near-constant $\Delta$ across all three difficulty levels (${\approx}+22.5$ and ${\approx}+24.8$\,pp respectively).
This is consistent with the scaffold supplying a coverage discipline (``write five focused tests, one invariant per function'') that benefits these models roughly equally regardless of bug difficulty.

\textbf{(c) Difficulty-amplified degradation.}
DeepSeek~V3.2 and Grok~4.1~Fast show negative overall $\Delta$, with greater degradation at higher difficulty (DeepSeek: $+2.7$/$-4.0$/$-7.8$\,pp; Grok: $-5.7$/$-5.2$/$-10.6$\,pp).
One possible explanation is strategy over-specification: the PBT template may constrain the input search in ways that are tolerable for L1 bugs but exclude the narrower trigger regions of L2--L3 bugs, consistent with the elevated search-related failure rates documented for these models in Appendix~\ref{app:failures-table}.
Gemini~3~Flash shows a weaker version at L3 only ($-0.4$\,pp, likely within noise) while retaining positive $\Delta$ at L1/L2 ($+8.8$/$+12.9$\,pp), making it a borderline case between patterns (b) and (c).

\textbf{(d) Large-$\Delta$-at-L1 pattern.}
Sonnet~4.6 shows its largest absolute $\Delta$ on L1 ($+13.8$\,pp), smaller at L2/L3 ($+5.7$/$+6.4$\,pp).
Sonnet~4.6 Baseline already achieves $79.3\%$ on L1 --- leaving limited headroom --- so the large L1 gain is notable; diminishing gains at L2/L3 may partly reflect that ceiling.
GLM~5.1 shows the same ordering ($+7.3$/$+3.7$/$+6.3$\,pp) but its Baseline L1 of $68.6\%$ is not near ceiling, so the explanation is less clear-cut.

We speculate that these patterns are consistent with two roles for PBT scaffolding: a \emph{framework bootstrapping} role (consistent with patterns b and d) that supplies the Hypothesis API and coverage discipline; and a \emph{constraint elicitation} role (consistent with pattern a) that substitutes for input-reasoning skills with diminishing returns as bug complexity grows.
The post-review ablation (Section~\ref{subsec:ablation}) partly separates the two roles: the framework instructions (Hypothesis plus a coverage goal) account for essentially all of the gain for Opus~4.8, whereas the taxonomy and template add a further $7$--$8$\,pp for Qwen~3.6 Plus and DeepSeek~V3.2 on problems with a test file ($12$--$13$\,pp on all problems), consistent with a constraint-elicitation role for mid-capability models. Why some models are harmed by the full scaffold remains open.

\subsection{Cross-model bug-finding complementarity}
\label{app:gals}

Not every hard bug is hard for every model. The three galois library
problems (\texttt{GALS-003}--\texttt{005}) illustrate this: each requires
deep algebraic domain knowledge (finite-field arithmetic, LFSR protocols,
polynomial factorization), yet model rankings are inconsistent across them.

\begin{itemize}[leftmargin=*, topsep=2pt, itemsep=1pt]
  \item \textbf{GALS-003} --- Sonnet Baseline achieves 100\% recall while
  several weaker models score $\leq 33\%$; yet DeepSeek PBT and Qwen 3.6
  PBT also reach 100\%.
  \item \textbf{GALS-004} --- Qwen 3.6 PBT leads at 83\%, while Sonnet PBT
  scores 67\% and DeepSeek PBT only 17\%. No model achieves full recall.
  \item \textbf{GALS-005} --- Gemini Baseline (92\%) and Qwen 3.6 PBT
  (92\%) lead; Sonnet Baseline scores 83\% but Sonnet PBT drops to 58\%.
\end{itemize}

This is not a general-capability inversion: Sonnet dominates overall
recall. Rather, it indicates that \benchname\ exposes problems for which
specific model architectures or scaffolded reasoning paths may be
differentially advantaged. For benchmark users, the actionable consequence
is that \emph{combining test files across models} is a recall-enhancing
strategy.

\subsection{Never-found bugs}
\label{app:never-found}

Two of 365 bugs are never reliably found (found on $\geq$2 of 3 runs) by any of the 16 (model, mode) cells; each is found in exactly one of three runs by a few cells.

\begin{itemize}[leftmargin=*, topsep=2pt, itemsep=1pt]
  \item \textbf{BIDC-004 bug\_3} (L2) --- the fast path of \texttt{BidictBase.\_update()} loses its ``target is empty'' guard, so calling \texttt{update()} on a non-empty bidict with another bidict as argument replaces the contents instead of merging them. Once a test makes that call, most generated inputs trigger the bug (about $70\%$ under default strategies, per \texttt{problem.yaml}); agents miss it because they rarely test \texttt{update()} with a bidict argument. It is found in one of three runs by Qwen3.5-35B-A3B PBT and Step~3.5 Flash PBT, and in the single default-effort Opus~4.8 Baseline run.
  \item \textbf{TMLK-005 bug\_1} (L3) --- when \texttt{Container.append} merges a super-table whose sections are out of order, it writes the merged value back to the first instead of the last body position. The trigger is a TOML document that splits one table into three or more sections interleaved with other tables, followed by a round trip; the output then contains duplicate table definitions. It is found in one of three runs by five original cells and by Opus~4.8 (PBT and framework-only).
\end{itemize}

The first is a coverage miss and the second a search miss; neither is found consistently by any model we evaluated.

\subsection{Failure-mode taxonomy (full)}
\label{app:failures-table}

To characterize \emph{how} agents fail, we sampled 20 failed
trajectories per (model, mode) cell (320 samples, uniform over problems)
and classified each into one of eight mutually-exclusive failure modes.
Classification is rule-based over the submitted \texttt{pbt\_test.py}
combined with per-function buggy/fixed pass patterns; a PBT-specific
refinement of the MAST taxonomy of \citet{pan2025mast} and the
trajectory-level analysis of \citet{merrill2026terminalbench}. Full
categorization rules and per-sample outputs are in
Appendix~\ref{app:failures} and the released
\texttt{paper/analysis/failure\_taxonomy.csv}.

\paragraph{Baseline vs PBT mode shows categorically different failure
profiles.}
In Baseline mode ($n=160$ classified failures), 59\% of failures are
\textbf{Incorrect Assertion} (the test's expected values are wrong ---
assertions fail against \emph{both} buggy and fixed libraries because
the agent fabricated incorrect outputs), 32\%
\textbf{Overly Concrete Test} (a few hardcoded inputs that happen not
to intersect the bug's trigger region), 8\% \textbf{Flaky Oracle}
(agent re-uses the library under test as its own oracle, typically via
a round-trip encode/decode whose asymmetric failure is masked), and
1\% \textbf{Under-Specified Property}. This pattern says Baseline
agents \emph{correctly construct tests} but aim them at the wrong
invariant.

In PBT mode ($n=160$), failure shifts to \emph{search pathologies}. The
single largest PBT-failure category is \textbf{Assume Misuse}
(31\% across PBT cells, peaking at 75\% for GLM 5.1, 50\% for DeepSeek
V3.2, and 40\% for Sonnet 4.6): the agent successfully designs a
\texttt{@given} test but calls \texttt{assume()} on a predicate that
happens to \emph{exclude the bug's trigger region} --- e.g.
\texttt{assume(val1 != val2)} in a test for a duplicate-value bug. The
Hypothesis engine then never samples the triggering input. A second
major PBT failure is \textbf{Wrong Strategy Range} (31\% across PBT
cells, peaking at 45\% for Qwen3.5-35B-A3B and 50\% for Gemini 3 Flash):
\texttt{st.integers()} where \texttt{st.integers(min\_value=32768,
max\_value=65535)} is needed, for example. In PBT mode, Baseline's
dominant Incorrect Assertion drops to 21\% (from 59\%) and Overly Concrete
Test vanishes ($0\%$), because the prompt's \texttt{@given} template
makes concrete-only tests syntactically unusual.

\emph{This diagnostic is specific to property-based-testing evaluation}: even when an agent correctly chooses
Hypothesis, its most common failure mode is filtering out the
counterexample region with \texttt{assume()}. Table~\ref{tab:failure-modes}
reports per-cell percentages; concrete examples are in
Appendix~\ref{app:failures}.

\begin{table}[h]
\centering
\small
\caption{Failure-mode distribution per (model, mode) cell, as percentages
of 20 sampled failures per cell. WS = Wrong Strategy Range, US =
Under-Specified Property, AM = Assume Misuse, OC = Overly Concrete Test,
FO = Flaky Oracle, SE = Setup Error, IA = Incorrect Assertion, Ot = Other.
Bold = row's largest category.}
\label{tab:failure-modes}
\begin{tabular}{llrrrrrrrr}
\toprule
Model & Mode & WS & US & AM & OC & FO & SE & IA & Ot \\
\midrule
Sonnet 4.6     & Baseline &  0 &  0 &  0 & 20 & 10 &  0 & \textbf{70} &  0 \\
Sonnet 4.6     & PBT      & 25 &  0 & \textbf{40} &  0 & 20 &  0 & 15 &  0 \\
GLM 5.1        & Baseline &  0 &  0 &  0 & 30 & 15 &  0 & \textbf{55} &  0 \\
GLM 5.1        & PBT      & 25 &  0 & \textbf{75} &  0 &  0 &  0 &  0 &  0 \\
Qwen 3.6 Plus  & Baseline &  0 &  0 &  0 & 25 &  0 &  0 & \textbf{75} &  0 \\
Qwen 3.6 Plus  & PBT      & 20 &  0 & \textbf{40} &  0 & 20 &  0 & 20 &  0 \\
DeepSeek V3.2  & Baseline &  0 &  5 &  0 & \textbf{45} & 10 &  0 & 40 &  0 \\
DeepSeek V3.2  & PBT      & 20 &  0 & \textbf{50} &  0 &  0 &  5 & 25 &  0 \\
Gemini 3 Flash & Baseline &  0 &  0 &  0 & 10 &  5 &  0 & \textbf{85} &  0 \\
Gemini 3 Flash & PBT      & \textbf{50} &  5 &  0 &  0 & 20 &  0 & 25 &  0 \\
Step 3.5 Flash & Baseline &  0 &  5 &  0 & \textbf{80} &  5 &  0 & 10 &  0 \\
Step 3.5 Flash & PBT      & \textbf{40} &  0 & 40 &  0 & 10 &  0 & 10 &  0 \\
Qwen3.5-35B-A3B   & Baseline &  0 &  0 &  0 & 45 &  5 &  0 & \textbf{50} &  0 \\
Qwen3.5-35B-A3B   & PBT      & \textbf{45} &  0 &  0 &  0 & 15 &  5 & 35 &  0 \\
Grok 4.1 Fast  & Baseline &  0 &  0 &  0 &  0 & 15 &  0 & \textbf{85} &  0 \\
Grok 4.1 Fast  & PBT      & 20 &  0 &  5 &  0 & 25 & 15 & \textbf{35} &  0 \\
\bottomrule
\end{tabular}
\end{table}

\paragraph{Implication for prompt design.}
The \textbf{Assume Misuse} finding is actionable: a PBT-scaffolding
prompt that warns explicitly \emph{``do not use \texttt{assume()} to
skip inputs that look suspicious; inputs that look suspicious are
exactly the ones that expose the bug''} is likely to recover a
substantial fraction of PBT-mode failures across most tested models. We
leave empirical verification to future work.

\subsection{Run-to-run stability (full)}
\label{app:stability}

All headline metrics are computed over $3$ independent runs per
(model, mode) cell and reported with $95\%$ percentile bootstrap
confidence intervals ($1{,}000$ resamples over (run, problem) instances).
Across the 16 original cells in Table~\ref{tab:headline}, the Recall CI
half-width ranges from $2.7$\,pp (Sonnet Baseline) to $4.9$\,pp (Step
PBT), with a median of $3.8$\,pp. Because runs of the same problem are
correlated, resampling problems instead (keeping each problem's runs together)
gives wider intervals (e.g.\ $\pm4.1$ instead of $\pm3.3$\,pp for Sonnet~4.6 PBT).
For the PBT$-$Baseline gap we therefore use a paired bootstrap over problems
($2{,}000$ resamples of the per-problem difference in mean recall), which accounts
for this correlation:
Qwen~3.6 Plus $+24.5$ $[+18.3,+30.4]$; Qwen3.5-35B-A3B $+22.9$ $[+17.3,+28.6]$;
Step~3.5 Flash $+20.3$ $[+12.3,+27.3]$; Sonnet~4.6 $+6.7$ $[+1.5,+11.4]$;
GLM~5.1 $+5.6$ $[-1.3,+12.2]$; Gemini~3 Flash $+6.6$ $[-1.6,+14.5]$;
DeepSeek~V3.2 $-3.2$ $[-10.1,+3.5]$; Grok~4.1 Fast $-8.1$ $[-15.0,-1.7]$
(exact differences; Table~\ref{tab:headline} shows differences of rounded values).

Run-to-run variance has two sources. Agents write different tests in different
runs: re-scoring the same Baseline test files reproduces 7,690 of 7,699 verdicts
(Section~\ref{subsec:ablation}), so Baseline variance comes almost entirely from
the agents. In PBT mode, Hypothesis's random search adds variance even for a fixed
test file. The \texttt{GALS-003} case
discussed in Appendix~\ref{app:gals} illustrates the latter: the
agent's strategy correctly targets polynomials in $\mathrm{GF}(q)$,
but only some random seeds generate one with the specific
$\geq 3$-distinct-degree-factor structure required to trigger the bug
within the default \texttt{max\_examples=200} budget. For problems with
known narrow triggers, raising \texttt{max\_examples} to $1{,}000$ or
higher would tighten the intervals further; a controlled study of this
budget parameter is left to future work.

\paragraph{Post-hoc prompt check.}

After the main runs, we wrote two paraphrased variants of each prompt
(\texttt{baseline\_v\{2,3\}.j2} and \texttt{pbt\_hypothesis\_v\{2,3\}.j2})
and ran Sonnet~4.6 on all \numproblems\ problems under each variant
(single run per variant) to check that headline recall is not an artifact
of a particular phrasing.

\emph{Baseline mode.}
Original (3-run mean): $0.765$; \texttt{v2}: $0.787$; \texttt{v3}: $0.650$.
The original and \texttt{v2} are within seed noise of each other
(the three seeds of the original prompt span $4.3$\,pp), while
\texttt{v3} is clearly worse.

\emph{PBT mode.}
Original (3-run mean): $0.834$; \texttt{v2}: $0.762$; \texttt{v3}: $0.842$.
The original and \texttt{v3} are within seed noise (the three seeds
of the original span $7.6$\,pp), while \texttt{v2} is clearly worse.

In both modes we kept the original prompt for the reported numbers.
We flag that Baseline~\texttt{v2} and PBT~\texttt{v3} are equally
defensible choices and release all candidates in \texttt{eval/prompts/};
a full sensitivity study (more models, more seeds per variant) is left
to future work.
The large \texttt{v3} Baseline drop ($-11.5$\,pp) is most plausibly attributable to an inferior paraphrase rather than fundamental instability, given that the concurrent \texttt{v2} variant deviates by only $+2.2$\,pp.

\subsection{Eval-phase timeout}
\label{app:eval-timeout}

One tuple (\texttt{glm51} $\times$ PBT $\times$ run~3 $\times$
\texttt{GALS-005}) hit the harness's eval-phase wall-clock limit.
The agent produced 96 \texttt{@given} test functions; during the
per-function F$\to$P scoring loop, test runs that reached the 30-second
command limit accumulated past the 1,800-second eval-phase budget. That trajectory's
four bugs are recorded as
\texttt{detection\_method=eval\_phase\_timeout} / \texttt{found=False}
rather than re-tried indefinitely. This is the sole abort in the
4,800-trajectory corpus. The harness now exposes an
\texttt{--eval-timeout} flag (default 1,800\,s) to make this failure
mode explicit for future runs.

\subsection{Compute resources and cost}
\label{app:compute}

All evaluations ran on a single 96-core, 512\,GB RAM CPU node (no GPU) with up to 20 concurrent Apptainer containers. Total wall-clock cost was approximately 200 node-hours. Aggregate API cost across the eight models was \$2{,}288, broken down as follows: Sonnet 4.6 PBT \$631; GLM 5.1 PBT \$364; Sonnet 4.6 Baseline \$308; all other cells $<$\$250 each.
The post-review evaluations ran on separate CPU servers with Docker and cost about \$1{,}750 in API fees through OpenRouter: Opus~4.8 on OpenHands \$1{,}126 (PBT, 4 runs: \$619; Baseline, 2 runs: \$113; framework-only, 3 runs: \$394); framework-only runs for Sonnet~4.6, Qwen~3.6 Plus, and DeepSeek~V3.2 \$245 in total; the two framework-only paraphrase runs \$105; and the three Claude Code runs \$274. The full research project additionally used compute for problem design, reference-model gating, pilot runs, and a discarded prompt variant (a ``hardened'' Baseline whose runs were dominated by missing test files); these are not reported.

\subsection{Post-review evaluations: setup, exclusions, and per-level results}
\label{app:post-review}

\paragraph{Setup.}
\emph{Opus~4.8 on OpenHands} uses the same SDK version, tool set, 200-iteration
cap, prompts, and scoring as the main evaluation, queried through OpenRouter with a
16,384-token output cap, on Docker rather than Apptainer containers built from the
same images. Some run configurations requested \texttt{reasoning\_effort=xhigh}
and others left it unset, but OpenHands SDK~1.11.5 forwards this parameter only for
model identifiers on its allow-list, which does not match
\texttt{anthropic/claude-opus-4.8}; no Opus~4.8 OpenHands request carried a
reasoning-effort setting, so all its runs share one configuration (provider
defaults) and are pooled: four PBT, two Baseline, and three framework-only runs.
The same allow-list also excludes the Sonnet~4.6 and Qwen~3.6 Plus identifiers, so
their original runs likewise used provider defaults.
The framework-only cells of Qwen~3.6 Plus, Sonnet~4.6, and DeepSeek~V3.2 also ran on
this Docker setup, months after these models' original cells.
\emph{Claude Code} runs headless (\texttt{claude -p}, permission prompts disabled)
with Opus~4.8 and Claude Code's own default settings, reached through a local
proxy that forwards to OpenRouter. The 15-minute limit ended 10 of 300 trajectories (\texttt{BIDC-004}
and \texttt{SYMP-003} in all three runs, \texttt{PRTI-002} in two, \texttt{PINT-003} and
\texttt{MRIT-004} in one); the median trajectory took 211\,s. Each problem gets a fresh container built from a clean Node/Python
image; the library is installed and the bug patches are applied and then removed
from the workspace before the agent starts. Network egress is then blocked with
\texttt{iptables} except for the model proxy, which we verify per instance, so the
agent cannot look up upstream fixes. The agent has 15 minutes and receives the
unchanged PBT prompt; its \texttt{pbt\_test.py} is scored by the same
two-container F$\to$P procedure.
\emph{Codex CLI} (GPT-5.5): in three pilot runs our integration did not produce a
test file, and we did not pursue it further.

\paragraph{Harness defect, excluded instances, and re-scoring.}
While preparing this version we found that the harness's exit handler stopped every
agent container on the host, not only its own. Under Apptainer, used for the 4,800
main trajectories, the handler did nothing, so those runs are unaffected. The
post-review OpenHands runs used Docker with several harness processes per host, so a
finishing process could stop the agents of the others mid-trajectory; we had reported
these as connection failures. We flag an OpenHands trajectory as stopped if it ended
without the agent's finish action within $[-5,+90]$\,s of another run's end or of a
cluster of at least three unfinished trajectories from at least two runs within
15\,s. This flags 106 of the 1,500 July OpenHands trajectories in the reported
cells, which we exclude and report as Excl.\ (Table~\ref{tab:post-review}). The
affected problems are not easier than average: their mean recall across the sixteen
original cells is $56.9\%$ versus $59.3\%$ for all problems, and the 106 instances
are spread over 71 problems; counting them as zero recall gives lower bounds (last
column). Because the defect could also stop scoring containers, and because an
unrelated fault had recorded Claude Code scoring errors (library images removed by a
host cache cleaner) as zero recall, we re-ran the F$\to$P scoring of all 2,000
post-review instances with the corrected harness; the agents' test files are
unchanged. Twelve trajectories of the two framework-only paraphrase runs were
interrupted while concurrent scoring processes on the host were restarted, and were
rerun. Re-scoring ran 40 instances concurrently. Every test function is limited to
30\,s, and a test that reaches the limit keeps running in its container while the
next one starts, so host load could turn a slow test into a failure. Re-scoring a
random sample of 48 instances that had at least one test reach the limit on an
otherwise idle host left 162 of 164 per-bug verdicts unchanged (one bug gained, one
lost). The harness now stops only its own containers, and
\texttt{paper/analysis/killed\_trajectories.csv} lists the flagged trajectories.

\begin{table}[h]
\centering
\small
\caption{Post-review cells. $n$: evaluated (run, problem) instances;
Excl.: trajectories excluded because the harness defect stopped the agent; No-file:
share of evaluated instances without a test file. Recall with $95\%$ bootstrap CI
half-width; the last column counts excluded instances as zero.}
\label{tab:post-review}
\setlength{\tabcolsep}{4pt}
\begin{tabular}{lrrrrrr}
\toprule
\textbf{Cell} & \textbf{Runs} & $n$ & \textbf{Excl.} & \textbf{No-file} & \textbf{Recall} & \textbf{Recall (excl.$=0$)} \\
\midrule
Opus 4.8 PBT                           & 4 & 377 & 23 & 1.6\%  & 81.2\%$\pm2.8$ & 76.5\% \\
Opus 4.8 Baseline                      & 2 & 200 & 0  & 0.0\%  & 39.3\%$\pm3.6$ & 39.3\% \\
Opus 4.8 framework-only                & 3 & 282 & 18 & 0.4\%  & 81.0\%$\pm3.1$ & 76.2\% \\
Claude Code + Opus 4.8 (PBT prompt)    & 3 & 300 & 0  & 1.3\%  & 78.6\%$\pm3.3$ & 78.6\% \\
Qwen 3.6 Plus framework-only           & 3 & 281 & 19 & 3.9\%  & 71.6\%$\pm3.8$ & 67.1\% \\
\quad paraphrase \texttt{v2}          & 1 & 100 & 0  & 2.0\%  & 76.1\%$\pm5.8$ & 76.1\% \\
\quad paraphrase \texttt{v3}          & 1 & 100 & 0  & 1.0\%  & 72.4\%$\pm6.6$ & 72.4\% \\
Sonnet 4.6 framework-only              & 1 & 78  & 22 & 7.7\%  & 79.3\%$\pm7.2$ & 61.8\% \\
DeepSeek V3.2 framework-only           & 2 & 176 & 24 & 21.6\% & 50.9\%$\pm6.2$ & 44.8\% \\
\bottomrule
\end{tabular}
\end{table}

\paragraph{Missing test files in the framework-only cells.}
For DeepSeek~V3.2, 32 of the 38 framework-only trajectories without a test file (22
of 26 in its first run) contain malformed file-creation tool calls by the model (the
file content argument is missing) that it never recovers from, a model tool-use
failure rather than a harness defect; in its original PBT runs only $12.0\%$ of
trajectories lacked a test file. Sonnet~4.6 lacks a test file in 6 of its 78
framework-only trajectories ($7.7\%$; original PBT runs: $4.7\%$). Because
trajectories without a file count as zero recall whereas stopped trajectories are
excluded, Table~\ref{tab:ablation} also reports the ablation on problems where every
condition produced a test file.

\paragraph{Framework-only prompt paraphrases.}
To check that the framework-only result does not hinge on its wording, we wrote two
paraphrases (\texttt{pbt\_framework\_only\_v\{2,3\}.j2}): \texttt{v2} rewords each
sentence and keeps the structure; \texttt{v3} is shorter and gives the same rules as a
numbered list. Both keep every instruction but drop the original's remark that the bug
lies in the scoped part of the API. With Qwen~3.6 Plus (one run each), recall over the
100 problems is $76.1\%$ (\texttt{v2}) and $72.4\%$ (\texttt{v3}) against $71.3\%$ for
the original prompt (runs: $72.8/72.9/69.2\%$); the paired differences, $+4.8$
[$-0.3,+10.0$] and $+1.1$ [$-4.7,+6.4$]\,pp, include zero. Both paraphrases elicit
more tests (median 31 and 43 test functions per problem, against 21.5).

\paragraph{Behaviour of Opus~4.8 under the Baseline prompt.}
Opus~4.8 writes a mean of 2.0 test functions per Baseline problem (median 2, in
both runs), versus means of 22.2 under the PBT prompt, 18.5 under the
framework-only prompt, and 24.1 in Claude Code. The original models write
2.8 (Grok~4.1 Fast) to 10.7 (GLM~5.1) test functions per Baseline problem.
Opus~4.8 finds at least one bug on $97\%$ of Baseline problems but all bugs on
$12\%$; its Baseline recall by level is $44.8/38.3/25.5\%$ (L1/L2/L3).

\paragraph{Recall by difficulty.}
Bug-level recall pooled over runs (as in Table~\ref{tab:recall-by-difficulty}),
L1/L2/L3/overall: Opus~4.8 PBT $87.6/82.8/75.6/82.1$; Opus~4.8 framework-only
$89.1/83.8/72.5/82.2$; Claude Code + Opus~4.8 $90.8/79.9/67.4/79.3$; Opus~4.8
Baseline $44.8/38.3/25.5/36.6$. Recall decreases monotonically with difficulty in
all four. Reliable coverage for Opus~4.8 uses its first three PBT runs; 38 bugs
have only two evaluated runs among them because of excluded instances, which makes
the criterion stricter for Opus.

\paragraph{Model-specific blind spots.}
Per-problem PBT recall of Opus~4.8 on OpenHands correlates more strongly with the
same model in Claude Code (Pearson $r=0.79$ over 100 problems) than with
Sonnet~4.6 ($r=0.41$). On four problems (\texttt{CONS-003}, \texttt{DCAC-003},
\texttt{PYCR-002}, \texttt{TRNS-001}) even Sonnet~4.6's worst run beats Opus~4.8's
best, and Claude Code's best run is also below Sonnet's worst run on all four. These misses belong to the model rather than the scaffold, in line
with the observation that no model dominates the hardest bugs.

\paragraph{Differences from figures reported during the review discussion.}
(i)~Figures reported then predate the harness defect described above: they used the
original scoring and, for Opus~4.8 PBT coverage and full recall ($90.7\%$ and
$58.7\%$), counted instances that failed for infrastructure reasons as failures; the
recall reported then ($81.1\%$) was the mean of three run means, and the Claude Code
figure ($80.3\%$) was a single run. Table~\ref{tab:post-review} re-scores every instance,
excludes stopped trajectories, and gives Claude Code three runs. The runs reported then as
``maximum'' and ``default'' reasoning effort did not differ in configuration (see
Setup above), so they carry no information about reasoning effort; this revision
pools them.
(ii)~The ablation reported then counted excluded framework-only instances as
written test files with zero recall, which lowered the framework-only column, and
it mixed Opus effort settings. Table~\ref{tab:ablation} corrects both; as a result,
the property taxonomy is not the dominant ingredient for Qwen~3.6 Plus or
Sonnet~4.6. (iii)~Two upstream pairings mentioned
then were imprecise: bidict~\#94 concerns deepcopy and pickle and is unrelated, and
transitions~\#175 corresponds to \texttt{TRNS-001} bug~4, not bug~1.
Table~\ref{tab:analogues} gives the verified mapping. (iv)~The blind-spot analysis
above uses a stricter criterion than the one reported then (seven problems).

\subsection{Scaling \benchname: feasibility and open challenges}
\label{app:scaling}

\benchname's size is limited by human review rather than by automation. Every
problem passes LLM-assisted bug proposal, author review against the design
checklist (Section~\ref{subsec:curation}), automated infrastructure checks, and a
reference-model gate that retires problems both reference models solve. The first
and third steps scale with compute; the second does not.

\paragraph{More Python libraries.}
The pipeline applies to any pure-Python library whose documentation states its
invariants and whose test suite is thorough enough to make the clean-pass
criterion meaningful. The main costs per library are building a cached container
image and curating agent-facing documentation. As models improve, the
reference-model gate should be re-run with stronger reference models to keep the
benchmark discriminative.

\paragraph{Other languages.}
The design needs only a PBT framework with composable generators and shrinking,
which exists in most major ecosystems (e.g.\ QuickCheck for Haskell, ScalaCheck
for Scala, jqwik for Java, fast-check for JavaScript/TypeScript, proptest for
Rust, PropEr for Erlang). The harness is language-agnostic at the level of
``apply patch, run test function, reverse patch, re-run''. We see four challenges.
(i)~For compiled languages, every fixed-bug-$N$ configuration requires a rebuild,
which multiplies evaluation cost by the number of bugs per problem unless builds
are incremental. (ii)~Per-function scoring needs a way to run test functions
individually with timeouts, which differs across test runners. (iii)~Documentation
quality and conventions vary widely, and a documentation-only oracle requires
documents that state invariants. (iv)~Stealth verification depends on an upstream
test suite that the injected bug must survive; weaker suites make the
clean-pass criterion less informative. Decontamination (canary strings,
periodically refreshed problems) applies unchanged.

\subsection{Real-world analogues of injected bugs}
\label{app:analogues}

Table~\ref{tab:analogues} lists fixed upstream defects that match injected bugs.
Each upstream item was checked against the issue tracker, and the tomlkit case
was reproduced: the reproducer from issue~\#148 fails on tomlkit~0.7.2, passes on
0.8.0 and on the unmodified benchmark version~0.14.0, and fails again once
\texttt{TMLK-005} bug~1 is applied. The matches cover specific bugs, not whole
problems; for the remaining bugs in these problems we found no verified upstream
counterpart. Independently of our bugs, property-based tests have exposed real
round-trip defects in two benchmark libraries (cbor2~\#163, a lost timezone sign;
dateutil~\#687, a parse/format round-trip failure for year~31). Conversely, a
recent dateutil change (PR~\#1581) fixed a Hypothesis property that was itself
wrong, illustrating the \emph{incorrect assertion} failure mode of
Section~\ref{subsec:diagnostics} in the wild.

\begin{table}[h]
\centering
\footnotesize
\caption{Verified upstream analogues of injected bugs, with the documented
property each violates.}
\label{tab:analogues}
\setlength{\tabcolsep}{3pt}
\begin{tabular}{p{1.9cm}p{3.5cm}p{3.0cm}p{4.6cm}}
\toprule
\textbf{Injected bug} & \textbf{Documented property} & \textbf{Upstream item} & \textbf{Relation} \\
\midrule
\texttt{TMLK-005} bug~1 & A table split into out-of-order sections parses to one table and round-trips. & tomlkit \#148 (2021), fixed in 0.8.0 & Same code branch: the bug changes the write-back index introduced by the fix and re-breaks the \#148 reproducer. \\
\texttt{TRNS-001} bug~4 & Machine-level \texttt{before} callbacks run before transition-level ones. & transitions \#175 / PR \#184 (0.5.0); PR \#370 (0.8.3) & Same mechanism: swaps the callback chains PR~\#184 ordered; \#370 fixed a real regression of this ordering in \texttt{AsyncMachine}. \\
\texttt{BIDC-004} bug~4 & \texttt{move\_to\_end(k, last=False)} moves $k$ to the front and keeps the others' order. & CPython bpo-25406 (2015) & Same defect (stale back-pointer when moving to the front) in \texttt{OrderedDict.move\_to\_end}, which \texttt{OrderedBidict} mirrors. \\
\texttt{BIDC-004} bugs~1--2 & \texttt{putall} is all-or-nothing: if it raises, the bidict and its inverse are unchanged. & bidict \#389 (2026), fixed in 0.24.0 & Same property, different mechanism; found upstream by a property-based test. \\
\bottomrule
\end{tabular}
\end{table}

\paragraph{Pre-existing upstream violations.}
Checking these analogues revealed that the pinned versions of two libraries
already violate a documented property: bidict~0.23.1 (\texttt{BIDC-004};
\texttt{putall} is not atomic for non-duplication exceptions, bidict~\#389) and
transitions~0.9.3 (\texttt{TRNS-001}; a parallel state's \texttt{on\_final} can
fire before all regions are final, transitions~\#715, fixed after 0.9.3).
The F$\to$P criterion never credits a test that detects only such a defect,
because the test fails in both the buggy and the fixed-bug configuration, so
scores remain valid; such tests do, however, cost agents effort on these two
problems. We document both cases in the dataset card.

%% file: checklist.tex
\section*{NeurIPS Paper Checklist}

\begin{enumerate}

\item {\bf Claims}
    \item[] Question: Do the main claims made in the abstract and introduction accurately reflect the paper's contributions and scope?
    \item[] Answer: \answerYes{}
    \item[] Justification: The abstract and Sections~\ref{sec:introduction}--\ref{sec:results} state four specific claims: (i) PBT-Bench measures a capability distinct from bug-repair, test-reproduction, and mutant-based test-generation benchmarks, and is positioned narrowly as a curated, documentation-grounded, Hypothesis-constrained benchmark (Section~\ref{sec:related_work}); (ii) the benchmark is not saturated, including by a newer frontier model and a commercial agent scaffold (Section~\ref{subsec:headline}); (iii) the gain from the PBT prompt is not monotone in model strength, and a controlled ablation attributes it for Opus~4.8 to the framework instructions rather than the property taxonomy, with the ablation's confounds stated (Section~\ref{subsec:ablation}); and (iv) no single configuration reliably covers every bug (Section~\ref{subsec:ensemble}). Each claim maps to a specific section, table, or figure.

\item {\bf Limitations}
    \item[] Question: Does the paper discuss the limitations of the work performed by the authors?
    \item[] Answer: \answerYes{}
    \item[] Justification: Section~\ref{sec:conclusion} discusses the limitations: the hand-curated bug distribution trades population fidelity for PBT-testability; the benchmark has 100 problems in one language (an extension roadmap is in Appendix~\ref{app:scaling}); most agents run on one scaffold, the frontier evidence comes from one model vendor, and Codex CLI was not evaluated; the ablation's framework arm bundles Hypothesis with a coverage goal, and three of its models ran later on different infrastructure; prompt sensitivity is only partially characterized; and the 200-example Hypothesis budget likely caps L2/L3 recall. Section~\ref{subsec:adversarial} and Appendix~\ref{app:leak-method} disclose a patch-backup leak in an early harness version with a null effect on recall, and Appendix~\ref{app:post-review} discloses a harness defect that stopped some post-review trajectories, which are excluded, and the re-scoring of all post-review instances with the corrected harness.

\item {\bf Theory assumptions and proofs}
    \item[] Question: For each theoretical result, does the paper provide the full set of assumptions and a complete (and correct) proof?
    \item[] Answer: \answerNA{}
    \item[] Justification: The paper contains no theoretical results.

    \item {\bf Experimental result reproducibility}
    \item[] Question: Does the paper fully disclose all the information needed to reproduce the main experimental results of the paper to the extent that it affects the main claims and/or conclusions of the paper (regardless of whether the code and data are provided or not)?
    \item[] Answer: \answerYes{}
    \item[] Justification: Section~\ref{sec:experimental_setup} specifies models, modes, agent scaffold (OpenHands SDK v1.11.5), max-iterations, worker counts, and the evaluation protocol; Appendix~\ref{app:post-review} specifies the post-review configurations (Opus~4.8, Claude Code, framework-only ablation). We release the full evaluation harness, all problem files, the 4{,}800 agent trajectories and the post-review trajectories, and per-run summaries. A single \texttt{bash scripts/run\_eval.sh} reproduces any cell on comparable hardware.

\item {\bf Open access to data and code}
    \item[] Question: Does the paper provide open access to the data and code, with sufficient instructions to faithfully reproduce the main experimental results, as described in supplemental material?
    \item[] Answer: \answerYes{}
    \item[] Justification: The complete benchmark (problems, docs, ground-truth), evaluation harness, trajectory corpus, and analysis scripts are released at a stable HuggingFace Datasets URL. Instructions are in the repository README.

\item {\bf Experimental setting/details}
    \item[] Question: Does the paper specify all the training and test details (e.g., data splits, hyperparameters, how they were chosen, type of optimizer) necessary to understand the results?
    \item[] Answer: \answerYes{}
    \item[] Justification: Section~\ref{sec:experimental_setup} specifies all hyperparameters (max-iterations 200; worker count 20; wall-clock cap 60 minutes per problem) and the post-review settings (provider-default reasoning settings for Opus~4.8 on OpenHands, Claude Code with its default settings and a 15-minute cap, Docker containers; Appendix~\ref{app:post-review}). Prompt templates are in \texttt{eval/prompts/} and also Appendix~\ref{app:prompts}.

\item {\bf Experiment statistical significance}
    \item[] Question: Does the paper report error bars suitably and correctly defined or other appropriate information about the statistical significance of the experiments?
    \item[] Answer: \answerYes{}
    \item[] Justification: Table~\ref{tab:headline} reports 95\% percentile bootstrap confidence intervals (1{,}000 resamples over (run, problem) instances) for all headline metrics; each original cell averages three independent runs, and the number of runs of each post-review cell is stated. PBT$-$Baseline gaps and the ablation effects (Table~\ref{tab:ablation}) use paired bootstrap CIs over problems (Appendix~\ref{app:stability}). Section~\ref{subsec:adversarial} reports a Wilcoxon signed-rank test with its p-value. Ensemble analysis in Section~\ref{subsec:ensemble} uses a majority-of-three reliable-coverage criterion.

\item {\bf Experiments compute resources}
    \item[] Question: For each experiment, does the paper provide sufficient information on the computer resources (type of compute workers, memory, time of execution) needed to reproduce the experiments?
    \item[] Answer: \answerYes{}
    \item[] Justification: Appendix~\ref{app:compute} reports compute: a 96-core 512\,GB CPU node with up to 20 concurrent Apptainer containers, approximately 200 node-hours total for the 4{,}800-trial evaluation, and an aggregate API cost of \$2{,}288 across the eight models; the post-review evaluations cost about \$1{,}750, and compute for problem design, pilot runs, and a discarded prompt variant is disclosed but not itemized.
    
\item {\bf Code of ethics}
    \item[] Question: Does the research conducted in the paper conform, in every respect, with the NeurIPS Code of Ethics \url{https://neurips.cc/public/EthicsGuidelines}?
    \item[] Answer: \answerYes{}
    \item[] Justification: The work involves no human subjects, no scraped personal data, and no dual-use risks beyond those present in general code-generation research. The benchmark itself is defensive (helps evaluate test-generation abilities).

\item {\bf Broader impacts}
    \item[] Question: Does the paper discuss both potential positive societal impacts and negative societal impacts of the work performed?
    \item[] Answer: \answerYes{}
    \item[] Justification: Positive: incentivizes LLMs that write more rigorous tests, raising software-quality standards. Negative: attackers could repurpose models trained on this benchmark to automatically discover exploitable vulnerabilities in deployed software. Mitigation: the benchmark tests only semantic invariants of pure-Python libraries (no network, no system calls), so the skills it develops are closer to defensive auditing than offensive exploitation.
    
\item {\bf Safeguards}
    \item[] Question: Does the paper describe safeguards that have been put in place for responsible release of data or models that have a high risk for misuse (e.g., pre-trained language models, image generators, or scraped datasets)?
    \item[] Answer: \answerNA{}
    \item[] Justification: The paper releases a benchmark (data and harness), not a trained model or generative asset. The benchmark itself poses no misuse risk.

\item {\bf Licenses for existing assets}
    \item[] Question: Are the creators or original owners of assets (e.g., code, data, models), used in the paper, properly credited and are the license and terms of use explicitly mentioned and properly respected?
    \item[] Answer: \answerYes{}
    \item[] Justification: Each library under test retains its original open-source license; bug patches, problem metadata, and harness code carry their own licenses (CC-BY-4.0 for data, MIT for code). Libraries are listed in Appendix~\ref{app:library-list}, and each problem's \texttt{problem.yaml} pins the library version. Hypothesis, OpenHands SDK, and other tooling are cited.

\item {\bf New assets}
    \item[] Question: Are new assets introduced in the paper well documented and is the documentation provided alongside the assets?
    \item[] Answer: \answerYes{}
    \item[] Justification: The benchmark is accompanied by a dataset card (\texttt{README.md} on HuggingFace), a Croissant metadata file (\texttt{croissant.json}), a canary-GUID declaration (\texttt{CANARY.md}) for training-corpus decontamination, and a repository with per-problem YAML metadata.

\item {\bf Crowdsourcing and research with human subjects}
    \item[] Question: For crowdsourcing experiments and research with human subjects, does the paper include the full text of instructions given to participants and screenshots, if applicable, as well as details about compensation (if any)? 
    \item[] Answer: \answerNA{}
    \item[] Justification: The paper involves no crowdsourcing or human-subject data collection. All bugs were authored by the research team.

\item {\bf Institutional review board (IRB) approvals or equivalent for research with human subjects}
    \item[] Question: Does the paper describe potential risks incurred by study participants, whether such risks were disclosed to the subjects, and whether Institutional Review Board (IRB) approvals (or an equivalent approval/review based on the requirements of your country or institution) were obtained?
    \item[] Answer: \answerNA{}
    \item[] Justification: The paper involves no human subjects.

\item {\bf Declaration of LLM usage}
    \item[] Question: Does the paper describe the usage of LLMs if it is an important, original, or non-standard component of the core methods in this research? Note that if the LLM is used only for writing, editing, or formatting purposes and does \emph{not} impact the core methodology, scientific rigor, or originality of the research, declaration is not required.
    \item[] Answer: \answerYes{}
    \item[] Justification: LLMs were used in three capacities. (i)~As evaluation targets (Section~\ref{sec:experimental_setup}). (ii)~As a design-assist tool that proposed candidate bug injections and patches; all candidates were reviewed by human authors against a design checklist before acceptance (Section~\ref{subsec:curation}). (iii)~As writing assistants for the manuscript; final content was reviewed and edited by the authors.

\end{enumerate}